\documentstyle[psfig]{lamuphys}

\def\amin{\ifmmode ^{\prime}\else$^{\prime}$\fi}
\def\asec{\ifmmode ^{\prime\prime}\else$^{\prime\prime}$\fi}
\def\45{RX\,J1239}       
\def\144{RX\,J1225}      
\def\0134{RX\,J0134}     
\def\7{RX\,J0119}     
\def\ros{{\sl ROSAT }}
\def\asca{{\sl ASCA}}

\def\ein{{\sl Einstein }}
\def\approxlt{\mathrel{\hbox{\rlap{\lower.55ex \hbox {$\sim$}}
        \kern-.3em \raise.4ex \hbox{$<$}}}}
\def\approxgt{\mathrel{\hbox{\rlap{\lower.55ex \hbox {$\sim$}}
        \kern-.3em \raise.4ex \hbox{$>$}}}}
\def\G{$\Gamma_{\rm x}$ }
\def\al{$\alpha_{\rm uv-x}$ } 

\begin{document}        

\title{\ros observations of warm absorbers in AGN}

\titlerunning{Warm absorbers in AGN}

\author{Stefanie Komossa, Henner Fink}

\institute{Max-Planck-Institut f\"ur extraterrestrische Physik, 85740 Garching, Germany}

\maketitle

\index{Komossa, S.|uu}
\index{Fink, H.}

\glossary{warm absorbers}
\glossary{AGN}
\glossary{narrow-line Seyfert 1 galaxies}   

\noindent{\bf Abstract:}

We study the properties of warm absorbers in several active galaxies (AGN) 
and present new candidates, using \ros X-ray observations.
Several aspects of the characteristics of warm absorbers are discussed: 
(i) constraints on the density and location of the ionized material 
are provided; 
(ii) the impact of the presence of the warm gas on other observable
spectral regions is investigated, particularly with respect to the 
possibility of a warm-absorber origin of one of the known high-ionization 
emission-line region in AGN; (iii) the possibility of dust mixed with the warm
material is critically assessed; and (iv) the thermal stability of the ionized gas is examined. 
Based on the properties of known warm absorbers, we then 
address the question of where else ionized absorbers might play a role in 
determining   
the traits of a class of objects or individual peculiar objects. 
In this context, the 
possibility to produce the steep soft X-ray spectra of narrow-line
Seyfert-1 galaxies by warm absorption is explored, as compared 
to the alternative of an accretion-produced soft excess.   
The potentiality of explaining the strong spectral variability
in {\0134}-42 and the drastic flux variability in NGC 5905
via warm absorption is scrutinized. 
 
\section{Introduction and motivation}

Absorption edges have been found in the X-ray spectra of several Seyfert galaxies
and are interpreted as the signature of ionized gas along the line of sight to the active nucleus.
This material, the so-called `warm absorber', provides an important
new diagnostic of the AGN central region,
and hitherto revealed its existence mainly in the soft X-ray spectral region.
The nature, physical state, and location of this ionized material,
and its relation to other nuclear components, ist still rather unclear.
E.g., an outflowing accretion disk wind,   
or a high-density component of the inner broad line region (BLR) 
have been suggested.    

The presence of an ionized absorber was first discovered in \ein observations    
of the quasar MR 2251-178 (Halpern 1984). 
With the availability of high quality soft X-ray spectra from \ros and \asca,
several more warm absorbers were found:    
they are seen in about 50\% of the well studied Seyfert
galaxies (e.g. Nandra \& Pounds 1992, Turner et al. 1993, Mihara et al. 1994, 
Weaver et al. 1994, Cappi et al. 1996) 
as well as in some quasars (e.g. Fiore et al. 1993, Ulrich-Demoulin \& Molendi 1996,
Schartel et al. 1997). 
More than one warm 
absorber imprints its presence on the soft X-ray spectrum of MCG-6-30-15 (Fabian 1996, Otani et al. 1996)
and NGC 3516 (Kriss et al. 1996).   
Evidence for an influence of the ionized material on non-X-ray parts of the spectrum  
is still rare:     
Mathur and collaborators (e.g. Mathur et al. 1994) combined 
UV and X-ray observations of some AGN to show
the X-ray and UV absorber to be the {\em same} component. 
Emission from NeVIII$\lambda$774, that may originate from a warm absorber,
 was discovered in several high-redshift quasars (Hamann et al. 1995).    

The warm material is thought 
to be photoionized by emission from the central continuum source in the 
active nucleus. Its degree of ionization is higher than that of the bulk of the BLR
and the gas temperature is typically of order 10$^5$ K.
The properties of hot photoionized gas were studied e.g. in Krolik \& Kallman (1984),
Netzer (1993), Krolik \& Kriss (1995), and Reynolds \& Fabian (1995). 

Here we analyze the X-ray spectra of several Seyfert galaxies, using
\ros PSPC observations that nicely trace the spectral region in which the warm absorption
features are located.
In a first part, the properties of the `safely' identified warm absorbers 
in the Seyfert galaxies NGC 4051, NGC 3227, and Mrk 1298  
are derived, as far as X-ray spectral fits allow.  
These absorbers span a wide range in ionization parameter. 
Their properties are then further discussed in light of
the known multi-wavelength characteristics of the individual Seyfert galaxies.   
E.g., the optical spectra show several features that may be directly
(high-ionization emission lines, strong reddening), or indirectly 
(FeII complexes) linked to the presence of the warm absorber.  
In a second part, we will assess whether some properties of other classes 
of objects or individuals can be understood in terms of warm absorption. 
In particular, the existence of objects with {\em deeper} absorption 
complexes that recover only beyond the \ros energy range is
expected (unless an as yet unknown fine-tuning mechanism is at work, 
that regulates the optical depths in the important metal ions).  
In this line, it is  examined 
whether the presence of a warm absorber
is a possible cause of the X-ray spectral steepness
in narrow-line Seyfert 1 (NLSy1) galaxies. 
Furthermore, we scrutinize whether the strong spectral variability in {\0134}-42, and the
drastic flux variability in NGC 5905
can be traced back to the influence of a warm absorber. 

\section{Models} 

The warm material was modeled using the photoionization code {\em Cloudy} (Ferland 1993).
We assume the absorber (i) to be photoionized by continuum emission of the central
pointlike nucleus, (ii) to be homogeneous,  
(iii) to have solar abundances (Grevesse \& Anders 1989),
and (iv) in those models that include dust mixed with the warm gas,
the dust properties 
are like those of the Galactic interstellar medium, 
and the abundances are depleted correspondingly (Ferland 1993).

The ionization state of the warm absorber can be characterized by the
hydrogen column density $N_{\rm w}$ of the ionized material and the
ionization parameter $U$, defined as
\begin {equation}
U=Q/(4\pi{r}^{2}n_{\rm H}c)
\end {equation}
where $Q$ is the
number rate of incident photons above the Lyman limit, $r$ is the distance between
central source and warm absorber, $n_{\rm H}$ is the hydrogen density
(fixed to 10$^{9.5}$ cm$^{-3}$ unless noted otherwise)
and $c$ the speed of light.  Both quantities, $N_{\rm w}$ and $U$, are determined from the X-ray
spectral fits.

The spectral energy distribution (SED) chosen for the modeling corresponds to
the observed multi-wavelength continuum in case of NGC 4051 and two of the NLSy1s, 
a mean Seyfert continuum with energy index $\alpha_{\rm uv-x}$ = --1.4 else. 
The influence of the non-X-ray parts of the continuum shape will be commented on in
Sect. 3.1, which will show the assumption of a mean Seyfert SED to be justified.   
However, the EUV part of the SED contributes to the numerical value of $U$. 
For better comparison of the ionization states of the discussed warm absorbers,
we also give the quantity $\tilde{U}$ for each object, which is $U$ normalized
to the mean Seyfert continuum with $\alpha_{\rm uv-x}$=--1.4.  

The X-ray data reduction was performed in a standard manner, using the 
EXSAS software package (Zimmermann et al. 1994). 

\section{NGC 4051}

\glossary{NGC 4051}
 NGC 4051 is a low-luminosity Seyfert galaxy with a redshift of z=0.0023,
classified as Seyfert~1.8 or NLSy1.
It is well known for its rapid X-ray variability and has been observed by  
all major X-ray missions.
Evidence for the presence of a warm absorber
was discovered by Pounds et al. (1994) in the \ros survey observation.
We have analyzed all archival observations of NGC 4051 (part of those were 
independently presented by McHardy et al., 1995) and PI data taken
in Nov. 1993.  
Emphazis is put on the latter observation, and the results refer to these data
if not stated otherwise (for details on the Nov. 93 observation see 
Komossa \& Fink 1997; for the earlier ones Komossa \& Fink 1996). 

\subsection{Spectral analysis}

A single powerlaw gives a poor fit to the X-ray spectrum of
NGC 4051 ($\chi{^{2}}_{\rm red}$ = 3.8) and 
the resulting
slope is rather steep (photon index $\Gamma_{\rm{x}}$ = --2.9). 
A warm absorber model
describes the spectrum well, with 
an ionization parameter of $\log U = 0.4$ ($\log \tilde{U}$ = 0.8), a large warm
column density of $\log N_{\rm w}$ = 22.67, and an intrinsic powerlaw spectrum
with index \G = --2.3. The unabsorbed (0.1-2.4 keV) luminosity is 
$L_{\rm x} = 9.5 \times 10^{41}$ erg/s (for a distance of 14 Mpc). 
The addition of
the absorber-intrinsic emission and far-side reflection
to the X-ray spectrum, calculated
for a covering factor of the warm material of 0.5, only negligibly changes
the results ($\log N_{\rm w}$ = 22.70) due to the weakness of these components.

We note, that the intrinsic, i.e. unabsorbed, powerlaw with index $\Gamma_{\rm x}$ = --2.3 is steeper
than the Seyfert-1 typical value with $\Gamma_{\rm x}$ = --1.9 and
it is also in its steepest observed state in NGC 4051.
We have verified that no additional
soft excess causes an apparent deviation from the canonical index, although
the presence of such a component, with $kT_{\rm bb} \simeq$ 0.1 keV, is found
in the high-state data. 

The observed multi-wavelength SED for NGC 4051 was used for the above
analysis, with $\alpha_{\rm uv-x}$=--0.7.   
In some further model sequences, the influence of the continuum shape
on the results was tested, by (i) 
the addition of an EUV black body component with variable temperature
and normalization (chosen such that there is no contribution of that 
component to the observed UV or X-ray part of the spectrum), 
and (ii) the inclusion of a strong IR spectral component as observed by IRAS
(although at least part of it is most probably due to emission by cold dust from
the surrounding galaxy; e.g. Ward et al. 1987). 
We find: (i) an additional
black body component in the EUV has negligible
influence on the X-ray absorption structure; 
(ii) an additional IR component strongly increases the free-free heating of the
gas and the electron temperature rises. The best-fit ionization parameter changes to 
\mbox{log $U$ = 0.2}.   

\begin{figure*} 
      \vbox{\psfig{figure=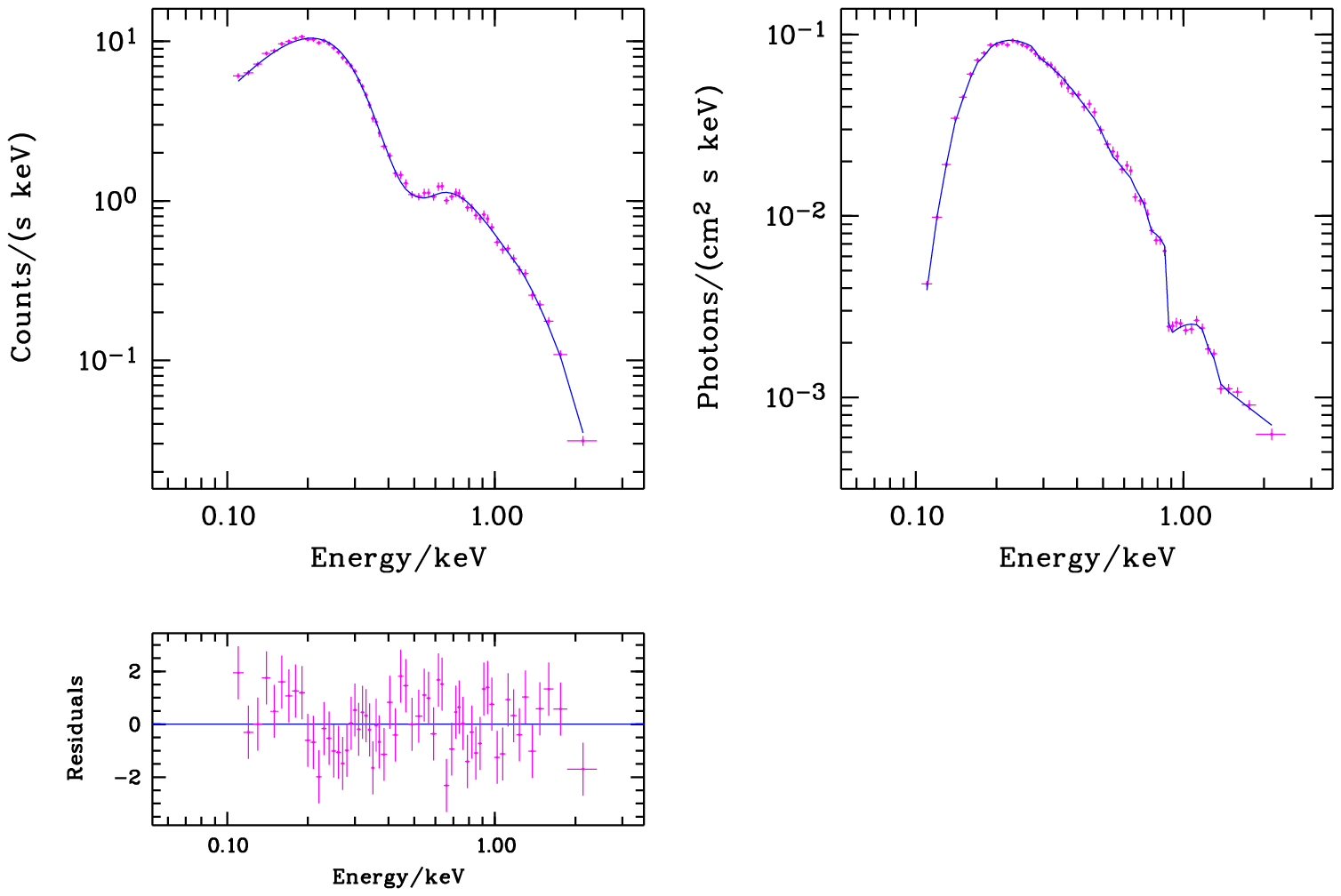,width=6.0cm,height=7.50cm,%
          bbllx=2.5cm,bblly=1.1cm,bburx=10.1cm,bbury=11.7cm,clip=}}\par
       \vspace*{-7.50cm}\hspace*{6.2cm}
      \vbox{\psfig{figure=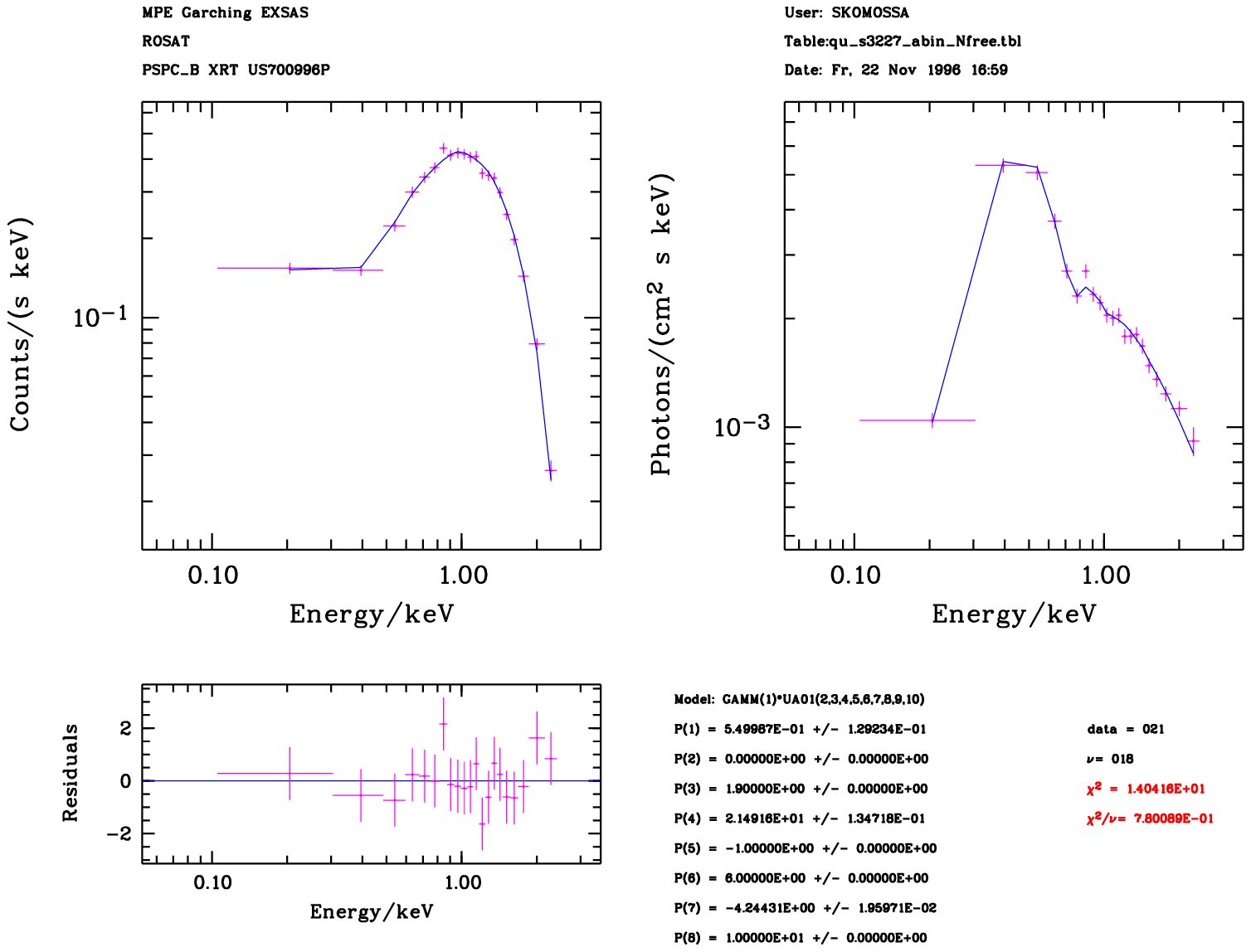,width=6.0cm,height=7.50cm,%
          bbllx=2.5cm,bblly=1.1cm,bburx=10.1cm,bbury=11.7cm,clip=}}\par
       \vspace{-0.5cm}
      \vbox{\psfig{figure=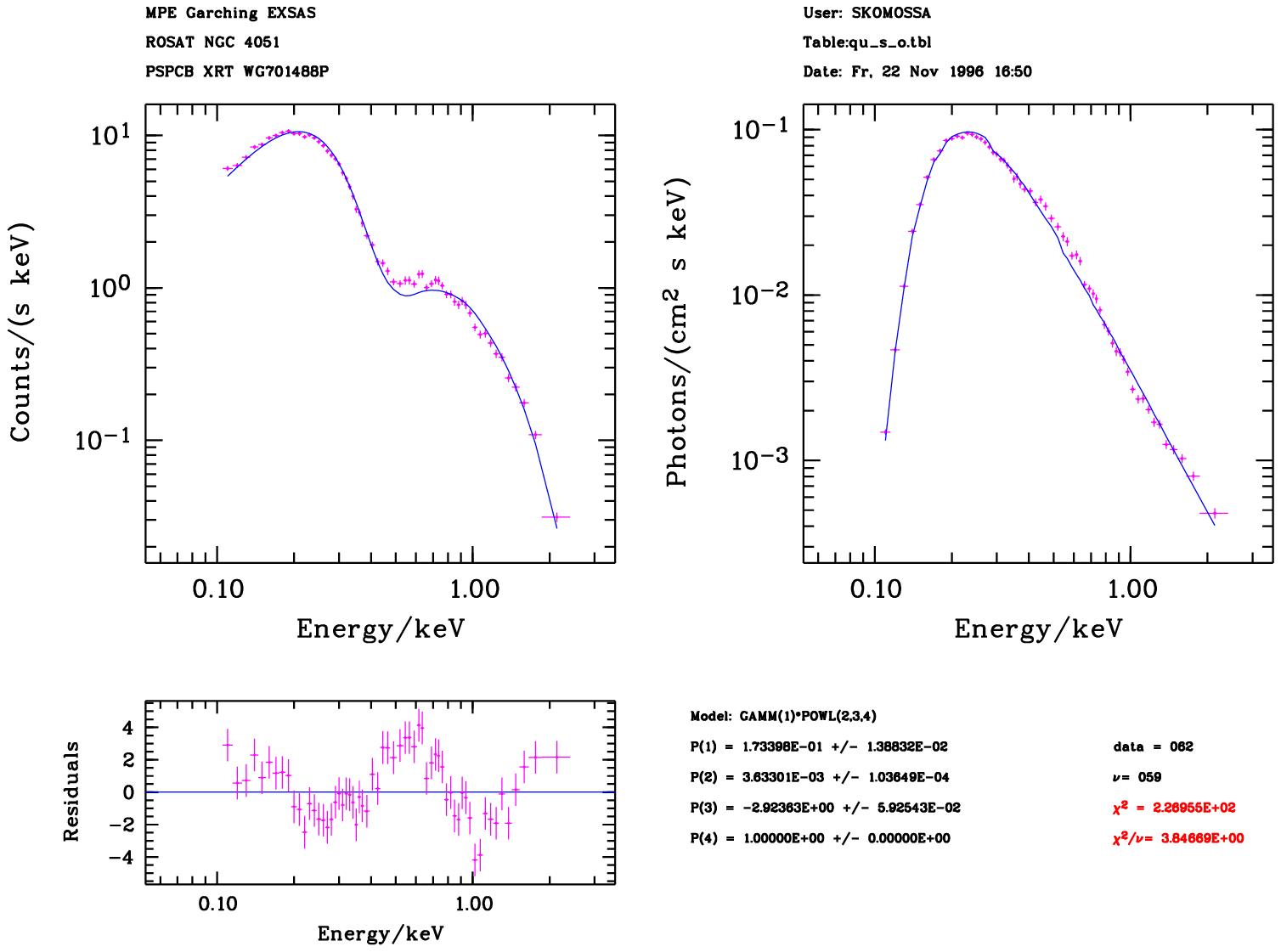,width=6.0cm,height=2.5cm,%
          bbllx=2.5cm,bblly=1.1cm,bburx=10.1cm,bbury=4.5cm,clip=}}\par
       \vspace*{-2.5cm}\hspace*{6.2cm}
      \vbox{\psfig{figure=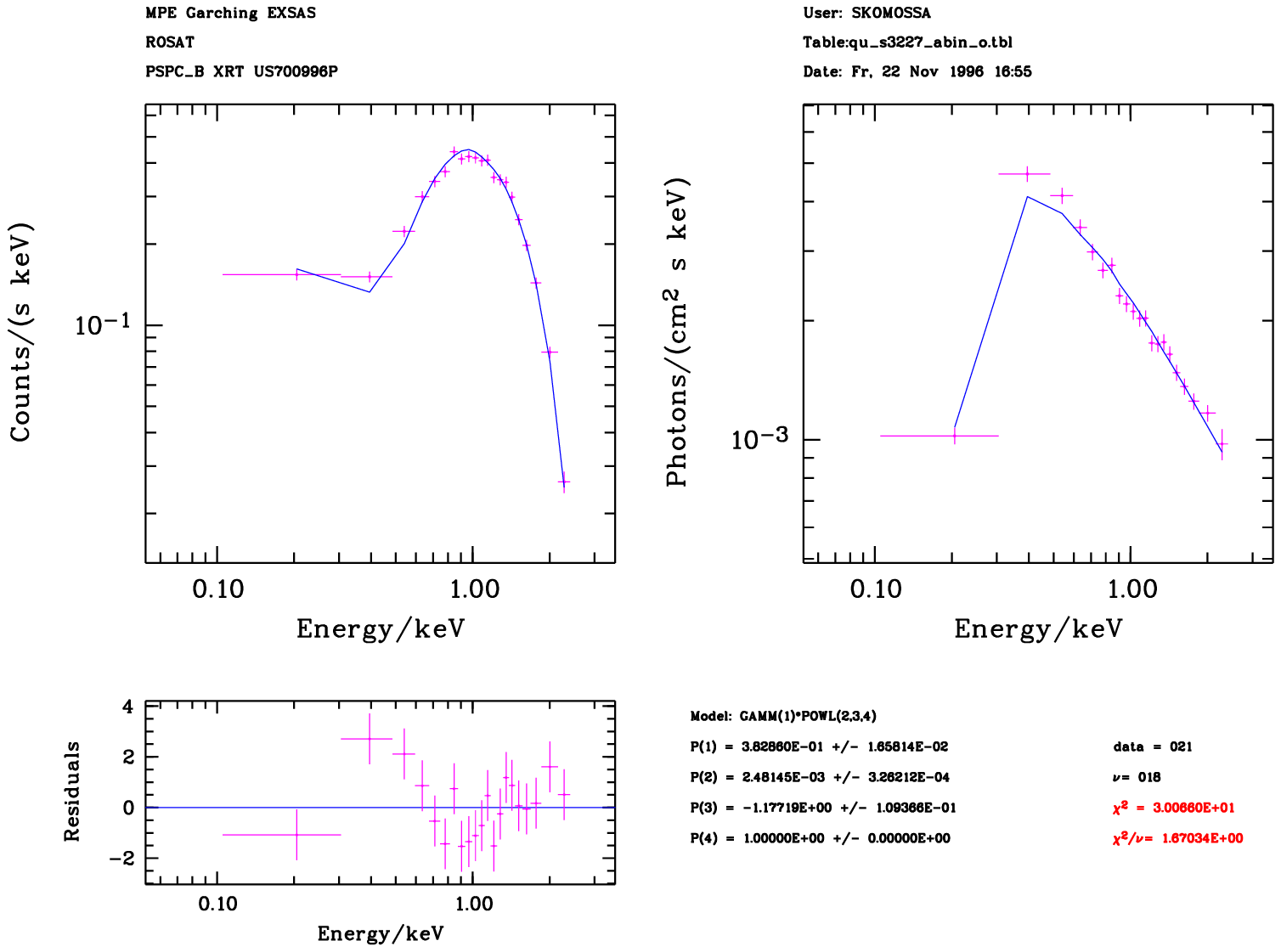,width=6.0cm,height=2.5cm,%
          bbllx=2.5cm,bblly=1.1cm,bburx=10.1cm,bbury=4.5cm,clip=}}\par
      \caption[]{Left: The upper panel shows the observed X-ray spectrum of NGC 4051 (crosses) and the
                best-fit warm absorber model (solid line). The next panel displays  
  the fit residuals for this model. For comparison, the residuals resulting from
  a single powerlaw description of the data are shown in the lower panel 
  (note the different scale in the ordinate). 
   Right: The same for NGC 3227.
}
      \label{SEDx}
\end{figure*}

\subsection{Temporal analysis}

\begin{figure*} 
      \vbox{\psfig{figure=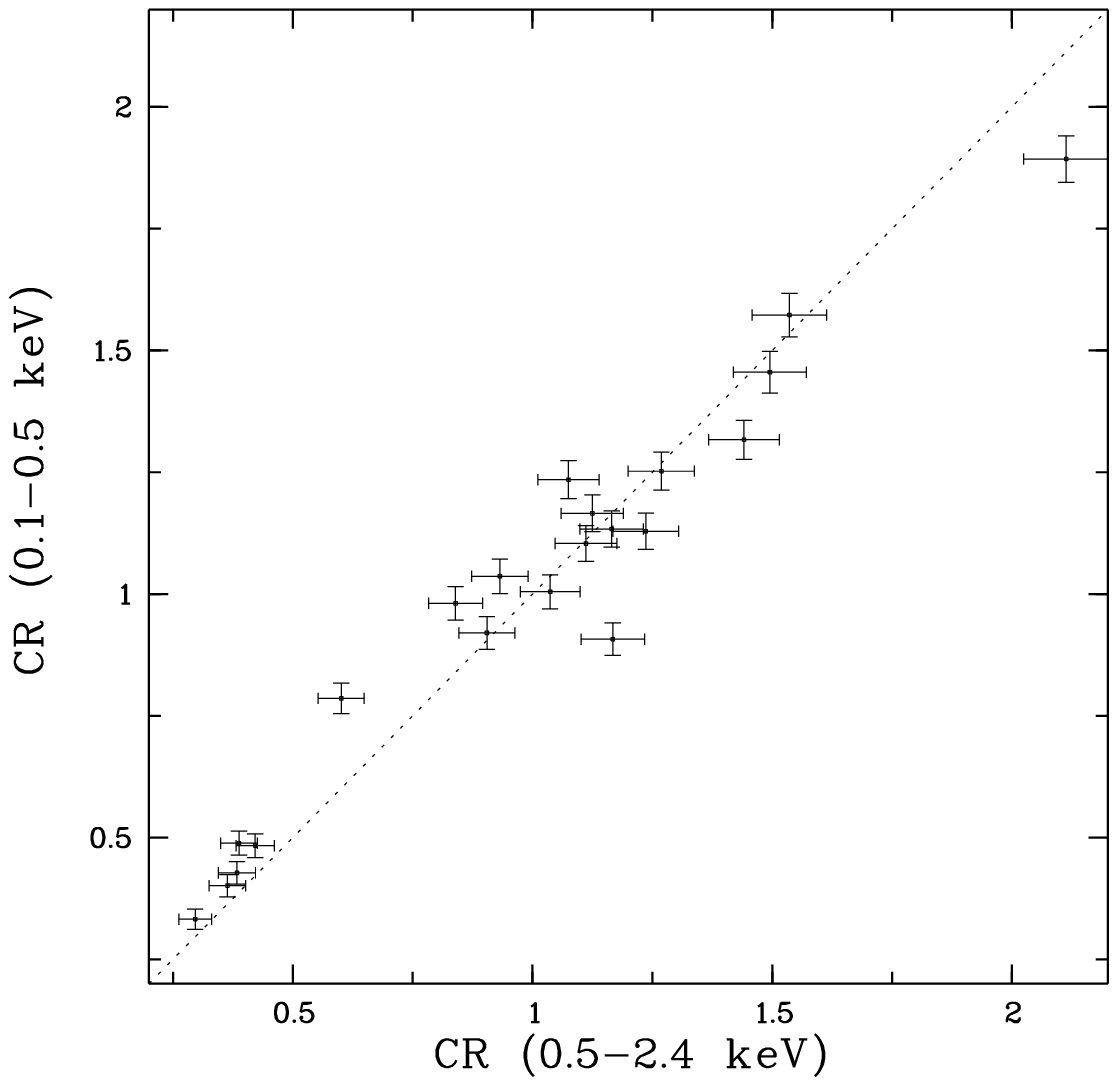,width=6.1cm,height=5.5cm,%
          bbllx=2.3cm,bblly=1.1cm,bburx=15.0cm,bbury=12.2cm,clip=}}\par
       \vspace*{-5.5cm}\hspace*{6.2cm}
      \vbox{\psfig{figure=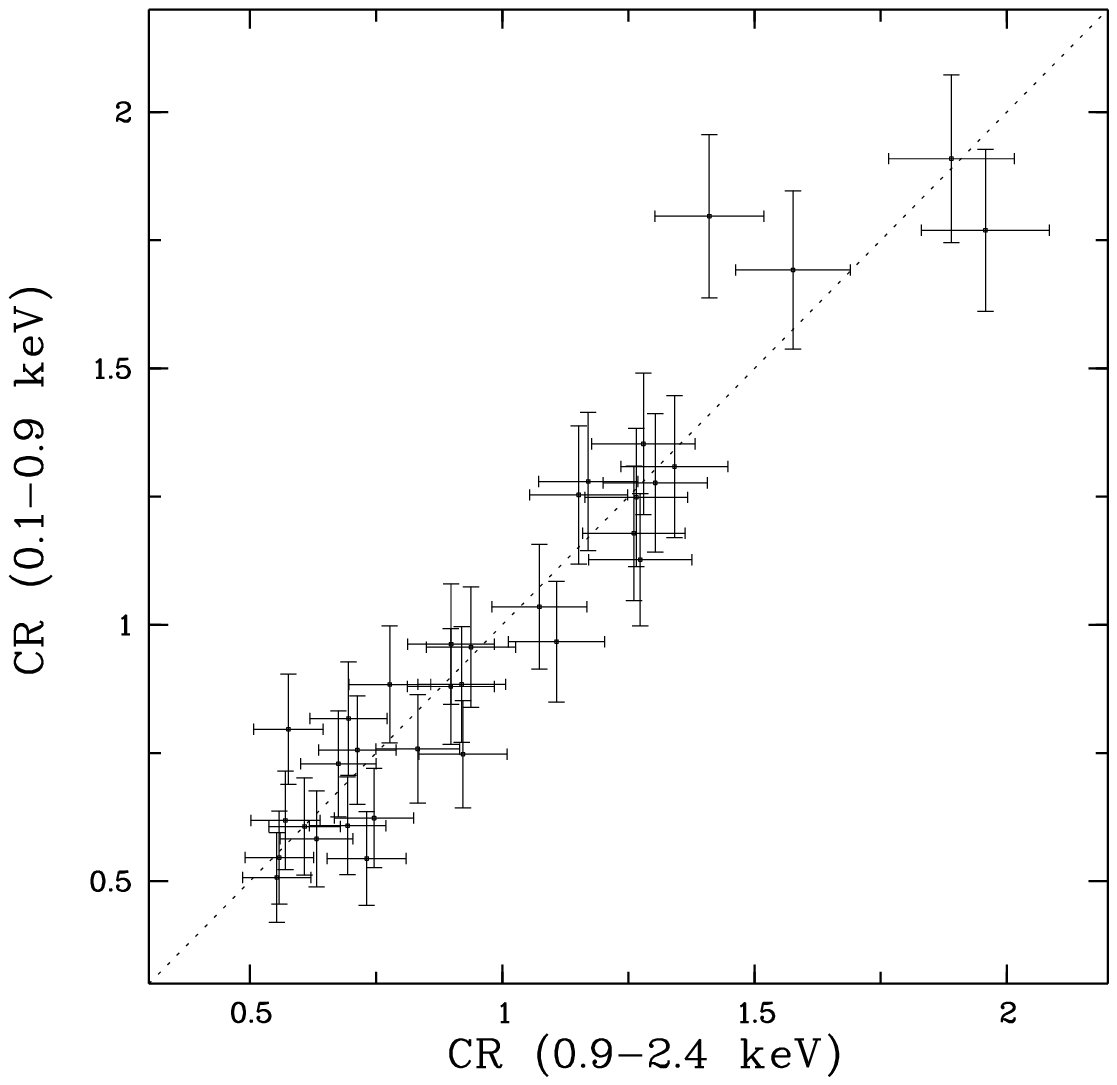,width=6.1cm,height=5.5cm,%
          bbllx=2.3cm,bblly=1.1cm,bburx=15.0cm,bbury=12.2cm,clip=}}\par
  \caption[]{Count rate CR in the soft \ros energy band versus 
count rate in the hard band, each normalized to the mean count rate
in the corresponding band. 
Left: NGC 4051, right: NGC 3227. The dotted line represents a linear correlation.  
}
      \label{sh}
\end{figure*}

We detect strong variability: The largest amplitude   
is a factor of $\sim$ 30 in count rate during the 2 year period of observations. 
In the Nov. 1993 observation, NGC 4051 is variable by about a factor of 6 within a day.
The amplitude of variability is essentially the same in the low energy ($E \le$ 0.5 keV;
dominated by the cold-absorbed powerlaw) and high energy (dominated by ionized-absorption)
region of the \ros band (Fig. \ref{sh}).
To check for variability of the
absorption edges in more detail, we have performed spectral fits to individual subsets of the
total observation (referred to as `orbits' hereafter). 
The best-fit ionization parameter turns out to be essentially constant
over the whole observation despite strong changes in intrinsic luminosity.
This constrains the, a priori unknown, density and hence
location of the warm absorber (see below).
%
\vspace{-1.0cm} 
\subsection{Properties of the warm absorber}

\subsubsection{Density} 
A limit on the density $n$ can be drawn from the spectral (non-)variability, i.e.
the constancy of $U$ 
during a factor of $>$ 4 change in intrinsic luminosity.
The reaction timescale $t_{\rm{rec}}$ of the ionized material 
is conservatively estimated from the lack of any reaction of the warm material
during the long low-state in orbit 7, resulting in a 
recombination timescale $t_{\rm{rec}} \approxgt$ 2000 s.
The upper limit on the density is given by
\begin{equation}
n_{\rm{e}} \approx {1\over t_{\rm{rec}}}~{n_{\rm{i}}\over n_{\rm{i+1}}}~{1\over A}({T\over 10^4})^{X}
\end{equation}
where $n_{\rm{i}}/n_{\rm{i+1}}$ is the ion abundance ratio of the
dominant coolant
and the last term is the
corresponding recombination rate coefficient $\alpha_{\rm{i+1,i}}^{-1}$ (Shull \& Van Steenberg 1982).
We find $n \approxlt$ $3 \times 10^{7}$cm$^{-3}$, dismissing a high-density BLR
component as possible identification of the warm absorber that dominates the Nov. 93 spectrum. 
The corresponding
thickness of the ionized material is $D \approxgt 2 \times 10^{15}$ cm.

\vspace{-0.5cm}
\subsubsection {Location} 

The location of the warm material is poorly constrained from X-ray spectral
fits alone.   
Using $Q = 1.6 \times 10^{52}$ s$^{-1}$ 
and the upper limit on the density,
yields a distance of the absorber
from the central power source of $r \approxgt 3 \times 10^{16}$ cm.
Conclusive results for the distance of the BLR in NGC 4051 from
reverberation mapping,
that would allow a judgement of the position of the warm absorber relative to the BLR, 
do not yet exist: Rosenblatt et al. (1992) find the continuum
to be variable with high amplitude, but no significant change in the H$\beta$ flux.

\vspace{-0.5cm}
\subsubsection{Warm-absorber intrinsic line emission and covering factor}

For 100\% covering of the warm material, the absorber-intrinsic H$\beta$
emission is
only about 1/220 of the observed $L_{\rm H\beta}$.
The strongest optical emission line predicted to arise from the ionized material 
is [FeXIV]$\lambda$5303. Its scaled intensity 
is [FeXIV]$_{\rm{wa}}$/H$\beta_{\rm{obs}}$ $\approx$ 0.01.  
This compares to the
observed upper limit of [FeXIV]/H$\beta \approxlt 0.1$ (Peterson et al. 1985)
and it is consistent with a covering of the warm material of less or equal 100\%.
Due to the low emissivity of the warm gas in NGC 4051,
no strong UV -- EUV emission lines are produced
(e.g. HeII$\lambda$1640$_{\rm{wa}}$/H$\beta_{\rm{obs}} \le 0.06$,
NeVIII$\lambda$774$_{\rm{wa}}$/H$\beta_{\rm{obs}} \le 0.2$). 
Consequently, no known emission-line component in NGC 4051
can be fully identified with the warm absorber. 
We have verified that this result holds independently of the exact value of density chosen,
as well as (IR or EUV) continuum shape.  

\vspace{-0.5cm}
\subsubsection{UV absorption lines} 

Due to the low column densities in the relevant ions, the predicted UV absorption
lines are rather weak.
The expected equivalent widths for the UV lines
CIV$\lambda$1549, NV$\lambda$1240 and Ly$\alpha$ 
are $\log W_{\lambda}$/$\lambda \simeq -4.0$,   
log $W_{\lambda}$/$\lambda \simeq -4.7$, and
$\log W_{\lambda}$/$\lambda \simeq -3.0$,  
respectively (for a velocity parameter $b$ = 60 km/s; Spitzer 1978).

\vspace{-0.5cm}
\subsubsection {Influence of dust}

Dust might be expected to survive in the warm absorber, depending on its distance
from the central energy source and the gas-dust interactions.
A dusty environment was originally suggested as an explanation for the
small broad emission-line widths in NLSy1 galaxies.   
Warm material with internal dust was proposed to exist in the quasar 
IRAS 13349+2438 (Brandt et al. 1996). 
The evaporation distance $r_{\rm ev}$
of dust due to heating by the radiation field is provided by 
$r_{\rm ev} \approx \sqrt{L_{46}}$ pc (e.g. Netzer 1990). 
For NGC 4051 we derive  $r_{\rm ev} \approx 0.02$ pc. 

Mixing dust of Galactic ISM properties (including both, graphite and
astronomical silicate; Ferland 1993) with the warm gas in NGC 4051
and self-consistently re-calculating the models
leads to maximum dust temperatures of 2200\,K (graphite) and 3100\,K (silicate), 
above the evaporation
temperatures (for a density of $n_{\rm H} = 5 \times 10^7$ cm$^{-3}$ and $U$, $N_{\rm w}$
of the former best-fit model).
For $n_{\rm H} \approxlt 10^6$ cm$^{-3}$, dust can survive throughout the absorber.
However, it changes the equilibrium conditions and
ionization structure of the gas. 
For relatively high ionization parameters,
dust very effectively competes with the gas
in the absorption of photons (e.g. Netzer \& Laor 1993).

\begin{figure*} 
      \vbox{\psfig{figure=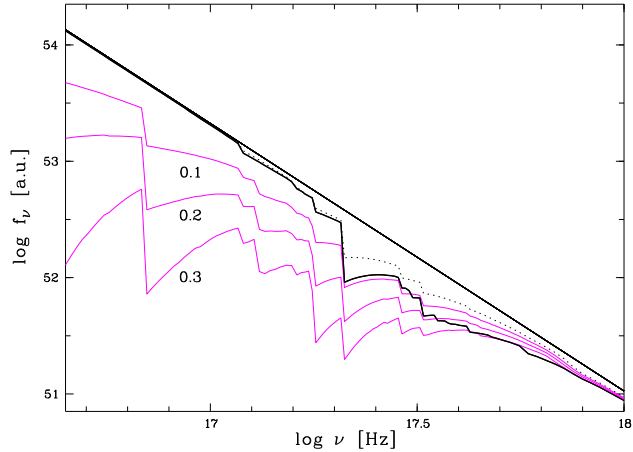,width=9.8cm,%
          bbllx=0.5cm,bblly=1.1cm,bburx=18.3cm,bbury=12.2cm,clip=}}\par
  \caption[]{Change in the X-ray absorption spectrum when dust is added to the warm absorber.
The thin straight line marks the intrinsic continuum, the fat line shows  
the best-fit warm-absorbed spectrum of NGC 4051. The dotted line corresponds to the same model,
except for depleted metal abundances. The thin solid lines represent models including dust (and
depleted abundances); the depletion factor of dust, relative to the standard Galactic-ISM
mixture, is marked.
}
  \label{dust_seq}
\end{figure*}

Firstly, re-running a large number of models, we find no successful fit of the X-ray spectrum.
This can be traced back to the relatively higher importance of edges from more lowly
ionized species, 
and particularly a very strong carbon edge (Fig. \ref{dust_seq}).   
There are various possibilities to change the properties of dust mixed with the warm material.
The one which weakens observable features, like the 2175 \AA ~absorption
and the 10$\mu$ IR silicon feature, and is UV gray, consists of
a modified grain size distribution, with
a dominance of larger grains (Laor \& Draine 1993). However, again, such models
do not fit the observed X-ray spectrum, even if silicate only is assumed to avoid a strong
carbon feature and its abundance is depleted by 1/10.

The absence of dust in the warm material would imply
either (i)  the {\em history} of the warm gas is such that dust was never able to
              form,  like in an inner-disc driven outflow (e.g. Murray et al. 1995,
               Witt et al. 1996)
    or     (ii) if dust originally existed in the absorber,  the {\em conditions in}
                the gas have to be such that dust is destroyed. In the latter 
                 case, one obtains an important constraint on the density (location) of
                 the warm gas, which then has to be high enough (near enough) to ensure
                 dust destruction. For the present case (and Galactic-ISM-like dust)
                 only a narrow range in density around
                 $n_{\rm H} \approx 5 \times 10^7$ cm$^{-3}$ is allowed. For lower densities,
                 dust can survive in at least part of the absorber and higher densities have already
                 been excluded on the basis of variability arguments.
%
%
\subsection{Evidence for an `EUV bump'}

The warm absorber fit shown in Fig. \ref{SEDx} shows some residual structure
between 0.1 and 0.3 keV that can be explained by an additional very soft
excess component.
Of course, the parameters of such a component are not well constrained from X-ray spectral
fits. One with $kT_{\rm bb}$ = 13 eV 
and an integrated absorption-corrected flux (between the Lyman limit and 2.4 keV)
of $F = 7 \times 10^{-11}$ erg/cm$^2$/s fits the data. It contributes about the same
amount to the ionizing luminosity as the powerlaw continuum used for the modeling
and has the properties of the EUV spectral component for which evidence is as follows: 

A lower limit
on the number rate $Q$ of ionizing photons in the unobserved EUV-part of the SED can be estimated
by a powerlaw interpolation between the
flux at 0.1 keV (from X-ray spectral fits) and the Lyman-limit (by extrapolating
the observed UV spectrum), which was used for the present modeling.
This gives $Q = 1.6 \times 10^{52}$ s$^{-1}$.

$Q$ can also be deduced from the observed H$\beta$ luminosity.
Both quantities are related via $Q = (2.1-3.8)\times$10$^{12}$ $L_{\rm H\beta}$ (Osterbrock 1989).
The mean observed $L_{\rm H\beta}$ =
7.8 $\times 10^{39}$ erg/s (Rosenblatt et al. 1992) yields
$Q = (1.6-2.5) \times 10^{52}$ s$^{-1}$. It is interesting to note that this is of the same order
as the lower limit determined from the assumption of a single powerlaw EUV-SED, suggesting
the existence of an additional EUV component in NGC 4051.

A third approach to $Q$ is via the ionization-parameter sensitive
emission-line ratio 
[OII]$\lambda$3727/[OIII]$\lambda$5007 (e.g. Schulz \& Komossa 1993), and using this method
we find
$Q \simeq (1.7 - 3.4) \times 10^{52}$ s$^{-1}$.

An additional EUV component may also explain
the observational trend that broad emission lines and observed continuum
seem to vary independently in NGC 4051
(e.g. Peterson et al. 1985, Rosenblatt et al. 1992),
as expected if the observed optical-UV continuum
variability is not fully representative of the EUV regime.

\section {NGC 3227}

\glossary{NGC 3227} 
NGC 3227 is a Seyfert 1.5 galaxy at a redshift of $z$=0.003.
Although studied in many spectral regions,
most attention has focussed on the optical wavelength range. NGC 3227
has been the target of
several high spectral and spatial resolution studies and BLR mapping campaigns. 
Ptak et al. (1994) found evidence for a warm absorber in an \asca~ observation.

\subsection {Spectral analysis}

For a single powerlaw 
description of the soft X-ray spectrum
we obtain a very {\em flat} spectrum ($\Gamma_{\rm{x}} \simeq$ --1.2), and strong systematic residuals remain.
A powerlaw with the canonical index of $\Gamma_{\rm{x}}$=--1.9, modified by
warm absorption, describes the data well.
The alternative description, a soft excess on top of a powerlaw, results
in an unrealistically flat slope with $\Gamma_{\rm{x}} \approx$ --1.1, in contradiction
to higher-energy observations (\G $\simeq$ --1.9, George et al. 1990; \G $\simeq$ --1.6, Ptak et al. 1994).

\subsection {Temporal analysis} 

We find the source to vary by a factor of 3.5 in count rate  
during the 11 days of the pointed observation.
The shortest resolved doubling timescale is 380 minutes.  
An analysis of the survey data reveals strong variability
between survey observation and pointing (which are separated
by $\sim$ 3y) with a  
maximum factor of
$\sim$ 15 change in count rate.  
Unfortunately, the low number of photons accumulated during the survey does not allow a 
detailed spectral analysis.

\begin{figure*} 
      \vbox{\psfig{figure=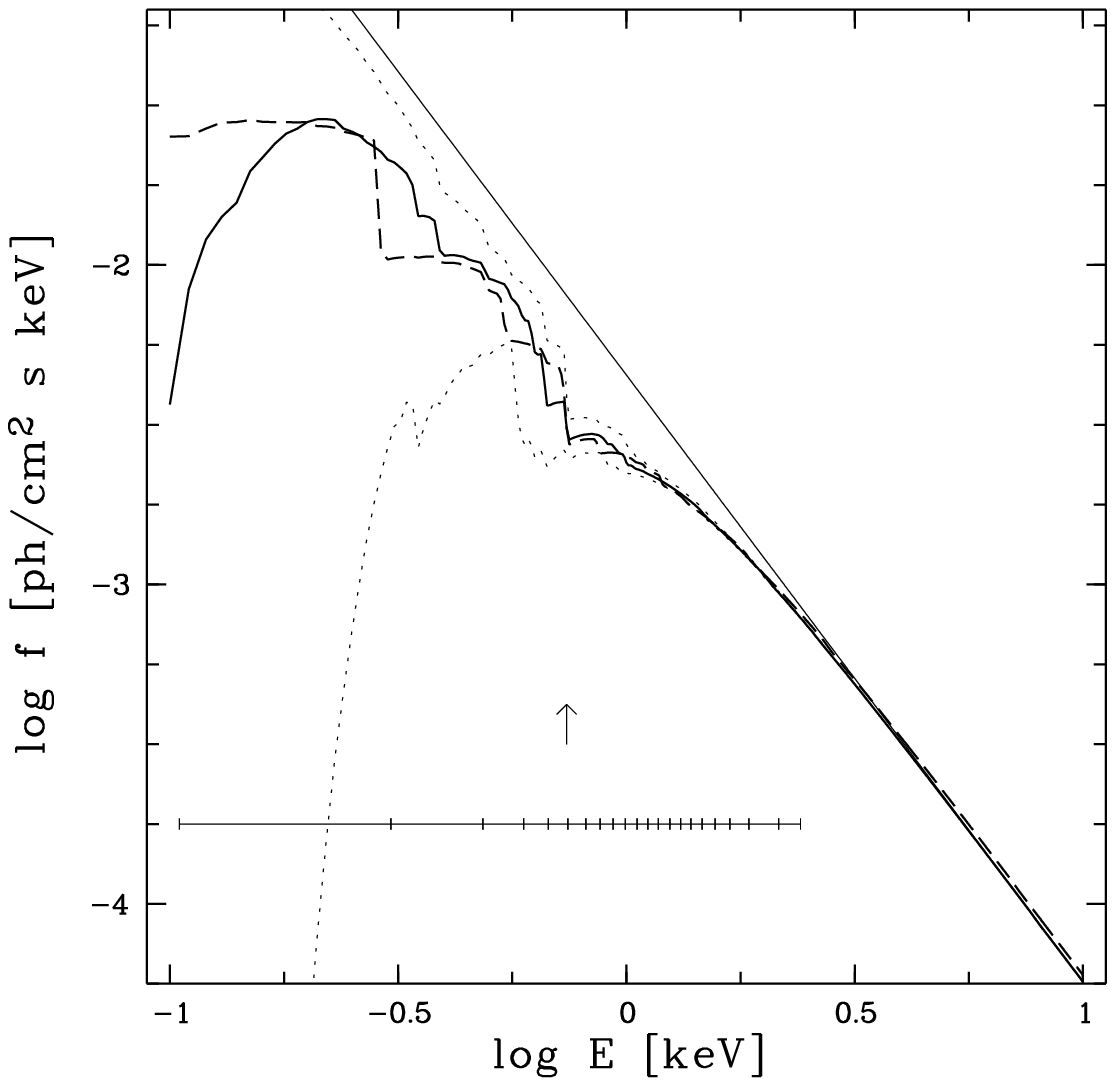,width=6.9cm,height=6.1cm,%
          bbllx=2.7cm,bblly=1.1cm,bburx=15.6cm,bbury=12.2cm,clip=}}\par
       \vspace*{-6.1cm}\hspace*{6.2cm}
      \vbox{\psfig{figure=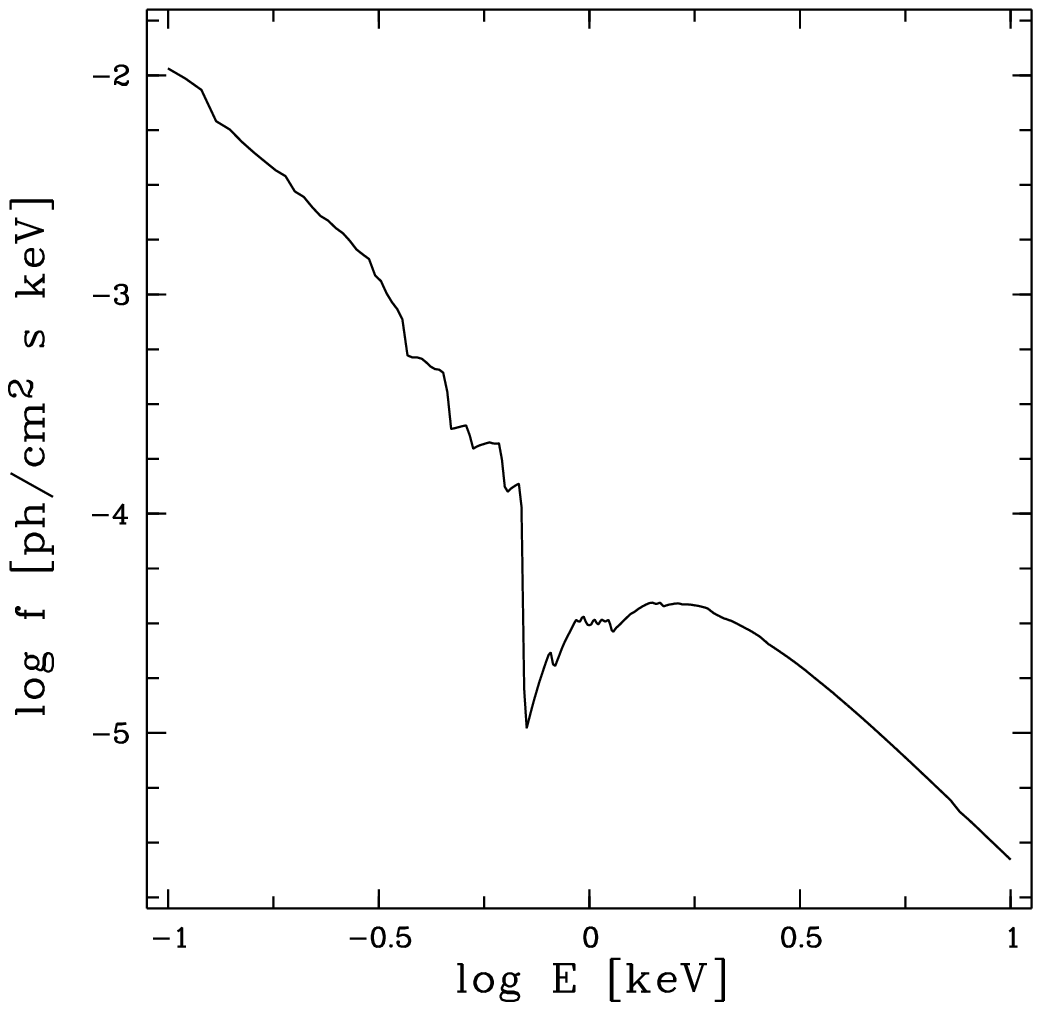,width=6.4cm,height=6.5cm,%
          bbllx=3.75cm,bblly=1.1cm,bburx=15.0cm,bbury=12.2cm,clip=}}\par
     \vspace{-0.4cm}
  \caption[]{Left: Warm-absorbed X-ray spectrum of NGC 3227 (solid line: dust-free, dashed: dusty model)
between 0.1 and 10 keV, corrected for cold absorption. The straight line corresponds
to the unabsorbed powerlaw. The dotted lines show the change in the absorption structure for
a factor of 2 (upper curve) and a factor of $1\over2$ change in luminosity, and consequently in $U$. The two curves
were  
shifted to match the normalization of the best-fit model, to allow a better comparison. 
The horizontal line at the bottom brackets the \ros sensitivity range, the vertical bars indicate
the bin sizes in energy that where used in the spectral fitting.
 The arrow marks the position of the OVII edge. Right: Warm-absorbed X-ray spectrum of Mrk 1298.
}
      \label{wa_def}
\end{figure*}

Dividing the data in a hard and soft band, according to a photon  energy
of larger or less than 0.9 keV, we find correlated variability between both bands
(Fig. \ref{sh}), implying that no drastic spectral changes have taken place. 
The soft band includes most of the warm absorption features 
(which extend throughout this whole
band; cf. Fig. \ref{wa_def})  
as well as the 
influence of the cold absorbing column, which consequently
cannot be disentangled from that of the ionized material. 
Spectral fits to individual `orbits'
of the total observation show the ionization parameter $U$
to be constant throughout the whole observation, independent of luminosity.
(In this study, the cold and warm column densities were fixed to the values determined for the total    
observation.) 

\subsection {Properties of the warm absorber}

\subsubsection {A {\em dusty} absorber?}
We find an ionization parameter of $\log U \approx -1.0$ and a warm
column density of $\log N_{\rm w} \approx 21.5$. The cold column,
$N_{\rm H} = 0.55 \times 10^{21}$ cm$^{-2}$ (corresponding to a reddening
of $A_{\rm v}$=0$^{\rm m}$.3, if accompanied by dust), is larger than the Galactic value
($N_{\rm H}^{Gal} \simeq 0.22 \times 10^{21}$ cm$^{-2}$), but not as large as
indicated by line reddening.  
The narrow H$\alpha$/H$\beta$ ratio measured by, e.g., Cohen (1983)
implies an extinction of
$A_{\rm v} \simeq$ 1.2.
At another time, Mundell et al. (1995a) report a much higher value,
$A_{\rm v} \simeq$ 4.5.
The intrinsic Balmer line ratios are expected to be close to the recombination value
under NLR conditions. 
A change in the H$\alpha$/H$\beta$ ratio has then to be attributed
to extinction by dust. However, dust {\em intrinsic} to the narrow line clouds
is not expected to vary by factors of several within years,
suggesting that the extinction is caused by an external absorber along the line
of sight.
If this dust is accompanied by an amount of gas as typically found in the
Galactic interstellar medium, a large cold absorbing column of $N_{\rm H} \simeq 8.3 
\times 10^{21}$ cm$^{-2}$ is expected to
show up in the soft X-ray spectrum. (We note that the present X-ray observation
and the optical observations by Mundell et al. (1995a) are not simultaneous,
but they are only separated by 6{$1\over2$} months.)
Such a large neutral column is {\em not} seen in soft X-rays. Dust mixed with
the warm gas could supply the reddening without implying a corresponding cold column.

Fitting a sequence of (Galactic-ISM-like) dusty absorbers provides
an excellent description of the soft X-ray spectrum, with $\log U \approx -0.25$ and
$\log N_{\rm w} \approx 21.8$.
The cold column is still slightly larger than the Galactic value, consistent with Mundell et al. (1995b),
who find evidence for HI absorption towards the continuum-nucleus of NGC 3227.

Another possibility to explain the comparatively low cold column observed in X-rays is
that we see scattered light originating from an extended component
outside the nucleus. However, in that case the X-ray emission should not be that rapidly
variable (e.g. Ptak et al. 1994; our Sect. 4.1).
An alternative that cannot be excluded, due to the non-simultaneity of the observations,
is that the material responsible for the reddening may have appeared only after the X-ray observation.
In this scenario, it may not be associated with the warm absorber at all, but consist of
thick blobs of dusty cold material crossing the line of sight.
Since this is a possibility,
we will include the dust-free warm absorber description of the data in the following discussion. 

%
\vspace{-0.4cm}
\subsubsection{Density and Location}

(i) For the dust-free best fit ($\log U \simeq -1.0$), the density-scaled distance of the warm
absorber is $r \simeq 10^{16}$cm $\times (n_{9.5})^{-0.5}$, which compares to the typical
BLR radius of $r \simeq 17$ ld = $4 \times 10^{16}$ cm, as determined from reverberation
mapping (Salamanca et al. 1994, Winge et al. 1995). 
(ii) For the dusty warm absorber, 
the gas density has to be less than about $10^{7}$ cm$^{-3}$, i.e. $r \ge 9 \times 10^{16}$ cm
$\times (n_{7})^{-0.5}$, to ensure dust survival.
Constant ionization parameter throughout the observed low-state in flux (Sect. 4.2) further implies
$n \approxlt 2 \times 10^6$ cm$^{-3}$; or even 
$n \approxlt 2.5 \times 10^{4}$ cm$^{-3}$, based on 
the assumption of constant $U$ during the total observation
(less certain, due to time gaps in the data), and still using $t_{\rm rec}$
as an estimate.    
The ionized material has to be located at least between BLR and NLR,
or even in the outer parts of the NLR, to extinct a non-negligible part of the NLR.  

\vspace{-0.4cm}
\subsubsection {Thermal stability} 
\begin{figure*}   
      \vbox{\psfig{figure=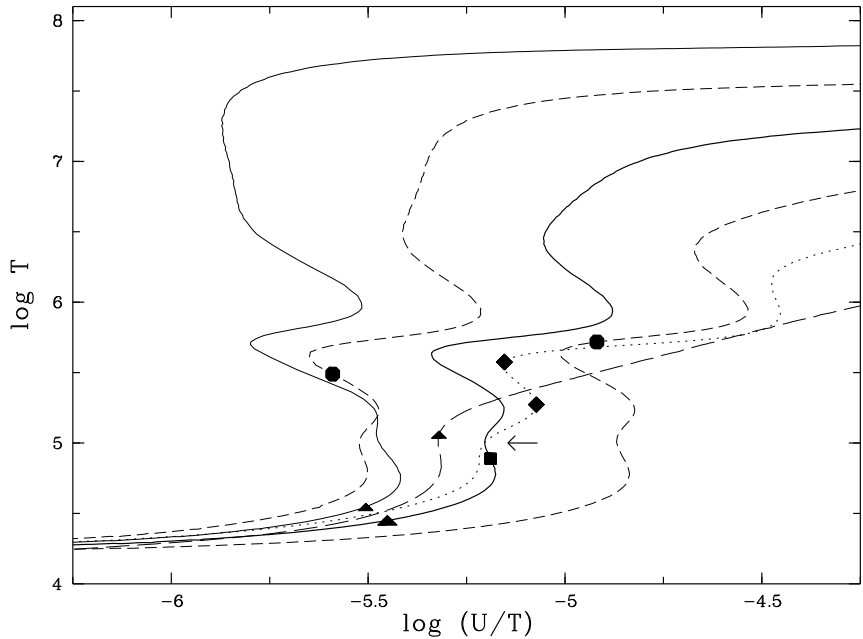,width=11.8cm,%
          bbllx=2.9cm,bblly=1.1cm,bburx=18.1cm,bbury=12.2cm,clip=}}\par
     \vspace{-0.1cm} 
  \caption[]{Equilibrium gas temperature $T$ versus $U/T$ and locations of warm absorbers.
The equilibrium curves are shown for different ionizing continua
and gas properties, and the positions of the warm absorbers discussed in the
present study are marked. The solid lines correspond to SEDs with (i) \G = --1.6 (left)
and (ii) \G = --1.9 (right), and \al = --1.4; long-dashed: the same continuum as (ii), 
but gas of depleted metal abundances and with dust; short-dashed: ionizing continua 
employed for the NLSy1 galaxies \45 (left, \al = --0.75) and \144 (\al = --1.9); 
dotted: observed SED of NGC 4051
with \G = --2.3. The symbols mark the positions that we derive for the warm absorbers
from X-ray spectral fits; triangles: NGC 3227 (for the dusty and dust-free absorber model);
square: Mrk 1298; lozenges: NGC 4051, Nov. 1991 observation (lower symbol) and Nov. 1993 
observation;   
arrow pointed to solid line: MCG-6-30-15 (Reynolds \& Fabian 1995);    
circles: \45 (left) and \144.     
}
 \vspace{-0.4cm} 
  \label{stab}
\end{figure*}

The thermal stability of the warm material is addressed 
in Fig. \ref{stab}.
Between the low-temperature ($T \sim 10^4$ K) and high-temperature ($T \sim 10^{7-8}$ K) 
branch of the equilibrium curve, there
is an intermediate region of multi-valued behavior of $T$ in dependence
of $U/T$, i.e. pressure (e.g., Guilbert et al. 1983; their Fig. 1).
It is this regime, where the warm absorber is expected to be located,
with its temperature typically found to lie around $T \sim 10^5$ K.
The detailed shape of the equilibrium curve in this regime depends strongly
on the shape of the ionizing continuum, and properties of the gas, e.g. metal abundances.
An analysis by Reynolds \& Fabian (1995) placed the warm absorber in
MCG-6-30-15 in a small stable regime within this intermediate region
(its  position is marked by an arrow in Fig. \ref{stab}).
The ionized material in NGC 3227 is characterized by a comparatively
low ionization parameter. Its location in the $T$ versus $U/T$ diagram is
shown for the models that provide a successful description
of the X-ray spectrum. The shape of the equilibrium curve is clearly modified
for the model including dust and correspondingly depleted metal abundances.

The locations of the warm absorbers in other objects of the present study
are also plotted. They are found to lie all around the intermediate stable region. 
Interesting behavior is shown by the warm material in two observations of NGC 4051.  
A detailed asessment of the thermal stability of ionized absorbers is still
difficult due to the strong influence of several parameters on the shape of the 
equilibrium curves. 
%
%
\vspace{-0.4cm}
\subsubsection {Line emission} 

From a sample of objects chosen to asses whether one of the
known emission-line regions in active galaxies can be identified with the warm absorber,
NGC 3227 represents the low-ionization end of the known warm absorbers. 
It also exhibits broad wings in H$\alpha$ that do not follow continuum variations
(Salamanca et al. 1994), indicative of a higher than usual ionized component,
that may be identified with the warm absorber.
(i) Dust-free warm absorber:  
The absorber-intrinsic luminosity in H$\beta$ is only about 1/30 of the broad observed H$\beta$
emission (Rosenblatt et al. 1994; de-reddened with $A_{\rm v}$=0.9). 
The scaled optical lines are weak,
e.g., [FeX]$_{\rm{wa}}$/H$\beta_{\rm{obs}}
 \approx$ 0.03. The strongest line in the ultraviolet is
CIV$\lambda$1549,~with 
\mbox{CIV$_{\rm{wa}}$/H$\beta_{\rm{obs}} \approx$~6}.
A comparison of the expected line strengths
with a typical BLR spectrum is shown in Fig. \ref{emi}.
(ii) In case of dust mixed with the ionized material, the overall
emissivity is reduced and no strong emission lines are predicted.  

\vspace{-0.5cm}
\subsubsection {UV absorption lines}

A comparison of UV absorption lines and X-ray absorption properties of
active galaxies was performed
by Ulrich (1988).
Her analysis of an IUE spectrum of NGC 3227 in the range
2000 -- 3200 \AA~revealed the presence of MgII$\lambda$2798 absorption
with an equivalent width of $\log W_{\lambda}$/$\lambda$ $\simeq$ --2.8.
The X-ray warm absorber in NGC 3227 is of comparatively low ionization parameter.
Nevertheless, the column density in MgII is very low, with $\log W_{\lambda}$/$\lambda \simeq$ --5.7
(and considerably lower for the dusty model).
Predictions of UV absorption lines that arise from the ionized material
are made for the more highly ionized species, e.g. $\log W_{\lambda}$/$\lambda \simeq$~--2.9 
(--~3.0)
 for CIV$\lambda$1549, and $\log W_{\lambda}$/$\lambda \simeq$ --3.0 
(--3.1) for NV$\lambda$1240 (for $b$ = 60 km/s).
The values in  
brackets refer to the warm absorber model that includes dust.

\section {Mrk 1298}

\glossary{Mrk 1298}
\glossary{PG 1126-041} 
Mrk 1298 (PG 1126-041) is a luminous Seyfert 1 galaxy at a redshift of $z$= 0.06.
The optical spectrum exhibits strong FeII emission. 
The presence of a warm absorber
in the X-ray spectrum was found by Wang et al. (1996).

\subsection {Spectral analysis}
Again, the deviations of the X-ray spectrum from a single powerlaw fit
are strong ($\chi{^{2}}_{\rm red}$=4.3)
with a rather steep slope of index $\Gamma_{\rm{x}} \simeq$ --2.6.
Among various spectral models compared with the data (including those consisting
of a soft excess on top of a powerlaw), only the warm absorber model provides a successful
description. Due to the rather low number of detected photons, the underlying
powerlaw index is not well constrained and was fixed to the canonical value of $\Gamma_{\rm{x}}$=--1.9.

\subsection {Properties of the warm absorber}
We find an ionization parameter of $\log U \simeq -0.3$ and a warm
column density of $\log N_{\rm w} \simeq$ 22.2.
The absorber-intrinsic line emission turns out to be negligible, with
$L_{\rm H\beta}^{\rm{wa}} = 1/740$ $L_{\rm H\beta}^{\rm{obs}}$,
except for OVI$\lambda1035_{\rm{wa}}$/H$\beta_{\rm{obs}} = 0.4$.
Wang et al. (1996) mention the existence of strong UV absorption lines in an IUE spectrum.
Here we find indeed rather large column densities in C$^{3+}$ and N$^{4+}$,
corresponding to an equivalent width for, e.g., CIV of $\log W_{\lambda}$/$\lambda \simeq -2.9$
(for $b$ = 60 km/s).
\vspace{-0.3cm} 
%
\begin{table}[ht]
  \caption{Properties of the warm absorbers in the three Seyfert galaxies. For comparison  
   results from a single powerlaw fit are shown. } 
  \begin{tabular}{llllllllll}
  \noalign{\smallskip}
  \hline
  \noalign{\smallskip}
    name & date of obs.~~~ & \multicolumn{5}{l}{warm absorber} & \multicolumn{3}{l}{single powerlaw}  \\ 
         & & $N_{\rm H}^{(1)}$ & \G~~~ & log $U$~~ & log $N_{\rm w}$~~ & $\chi^2_{\rm red}$~~~~~ & 
                                                               $N_{\rm H}^{(1)}$ & \G~~~ & $\chi^2_{\rm red}$ \\  
  \noalign{\smallskip}
  \hline
  \noalign{\smallskip}
    NGC 4051~~ & Nov. 1991 & 0.13$^{(2)}$ & --2.2 & ~~0.2 & 22.5 & 1.1 & 0.18 & --2.8 & 2.8 \\
  \noalign{\smallskip}
             & Nov. 1993 & 0.13   & --2.3 & ~~0.4  & 22.7 & 1.1 & 0.17 & --2.9 & 3.8 \\            
  \noalign{\smallskip}
  \hline
  \noalign{\smallskip}
    NGC 3227~~ & May 1993 & 0.55 & --1.9 & --1.0   & 21.5 & 0.8 & 0.38 & --1.2 & 1.7 \\
  \noalign{\smallskip}
             &  & 0.55 & --1.9 & --0.3 & 21.8 & 0.7 $^{(3)}$ & & & \\
  \noalign{\smallskip}
  \hline
  \noalign{\smallskip}
    Mrk 1298~~ & June 1992 & 0.44$^{(2)}$ & --1.9 & --0.3 & 22.2 & 0.9 & 0.44$^{(2)}$ & --2.6 & 4.3 \\   
  \noalign{\smallskip}
  \hline
  \noalign{\smallskip}
     \end{tabular}
  \label{tab1}

  \noindent{\small
  $^{(1)}$ in 10$^{21}$ cm$^{-2}$ ~ $^{(2)}$ Galactic value ~ $^{(3)}$ model including dust}
\end{table}
%
\section { Narrow Line Seyfert 1 Galaxies }

Several extremely X-ray soft Seyferts have been found recently that belong
to the class of NLSy1 galaxies (e.g. Puchnarewicz et al. 1992), which are characterized by
narrow Balmer lines, and often strong FeII emission (e.g. Goodrich 1989).
One interpretation of the X-ray spectral steepness is the dominance
of the hot tail of emission from an accretion disk (e.g. Boller et al. 1996). 
Another one, which will be explored below, is the presence of a warm absorber.  

In fact, a natural consequence of the observation of absorption {\em edges} in the 
X-ray spectra of Seyfert galaxies, is to also expect objects with deeper absorption, with
mainly the `down-turning' part of the absorption-complex
being visible in the \ros sensitivity region.
Good such candidates are ultrasoft AGN, and some are studied in the following.

\subsection {RX\,J1239.3+2431 and RX\,J1225.7+2055} 
\glossary{RX\,J1239.3+2431}
\glossary{RX\,J1225.7+2055} 
The 2 NLSy1 galaxies were
discovered in \ros survey observations
and exhibit very steep X-ray spectra (formally $\Gamma_{\rm{x}} \simeq $--3.7 and --4.3;
Greiner et al. 1996) and there is
slight evidence for a spectral upturn within the \ros band.
A description of the X-ray spectra in terms of warm absorption is found to be successful,
albeit not unique
due to the poor photon statistics of the survey data.
We find $\log U \approx$ 0.8 ($\log \tilde{U} \approx$ 0.4) and
$\log N_{\rm w} \approx$ 23.2 for \144, and  $\log U \approx$ --0.1
($\log \tilde{U} \approx$ 0.2) and
$\log N_{\rm w} \approx$ 22.8 for \45.

\begin{figure*}[htb]
      \vbox{\psfig{figure=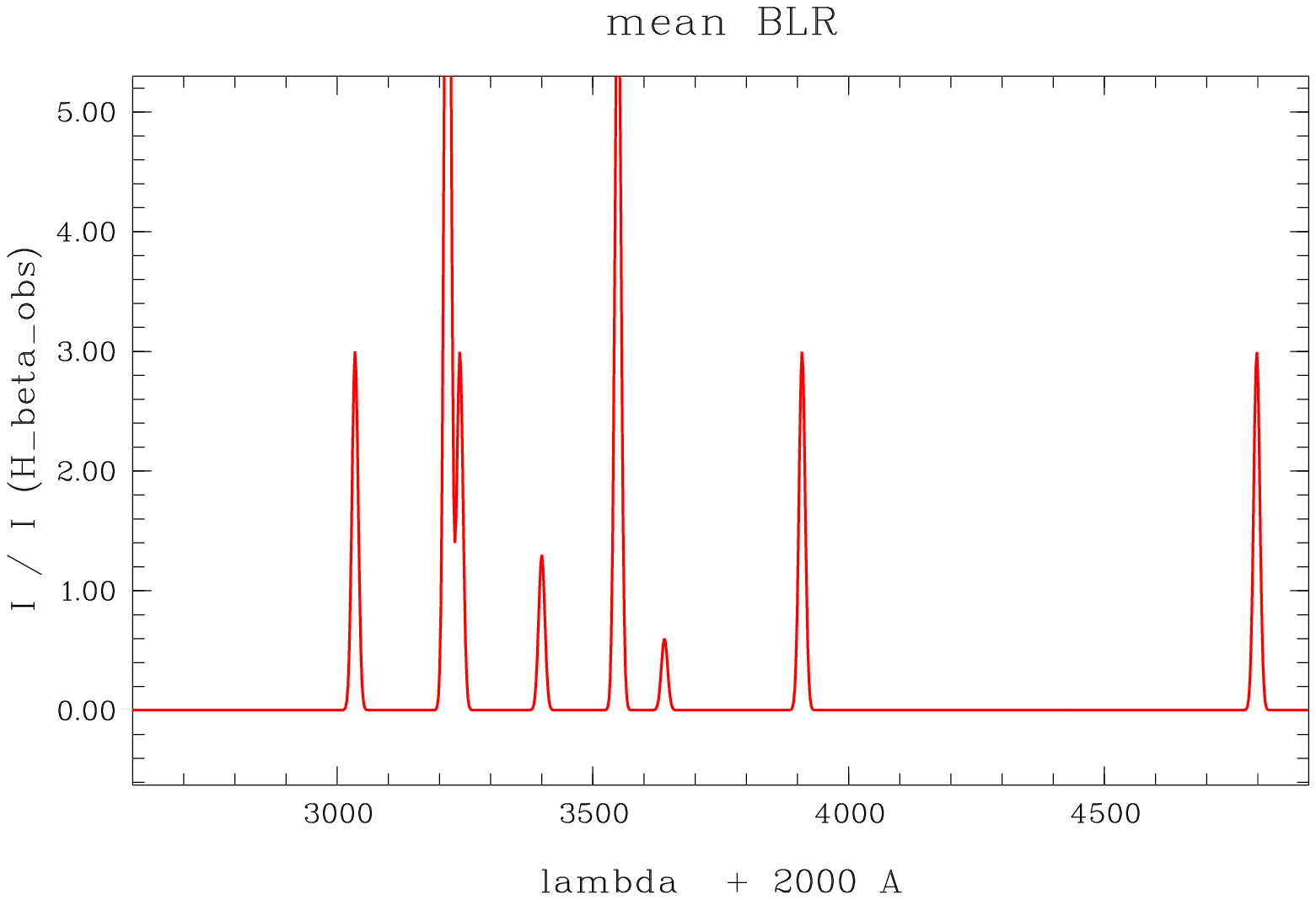,width=6.0cm,height=4.0cm,%
          bbllx=2.3cm,bblly=1.9cm,bburx=18.4cm,bbury=12.3cm,clip=}}\par
       \vspace*{-4.0cm}\hspace*{6.2cm}
      \vbox{\psfig{figure=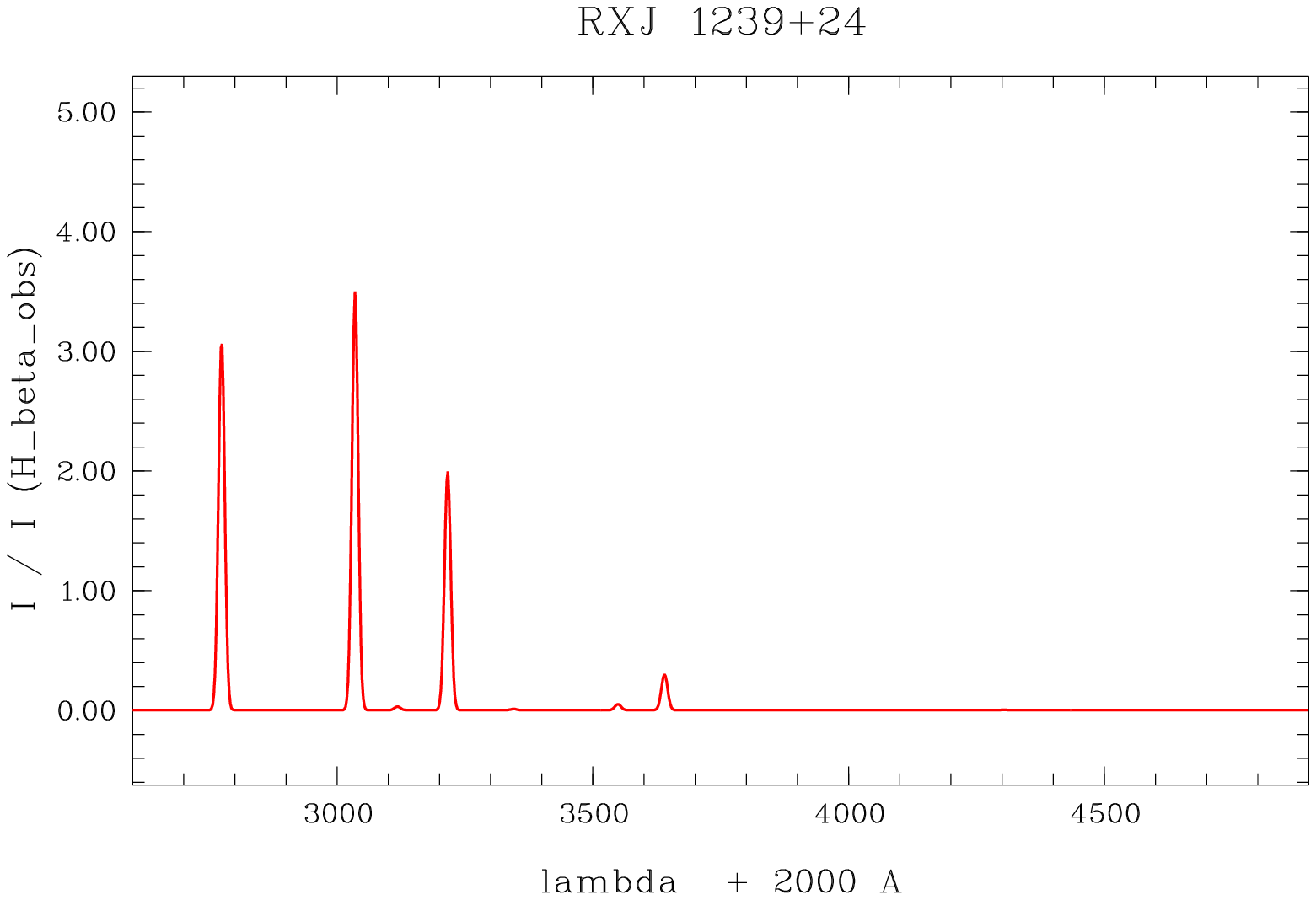,width=6.0cm,height=4.0cm,%
          bbllx=2.3cm,bblly=1.9cm,bburx=18.4cm,bbury=12.3cm,clip=}}\par
      \vbox{\psfig{figure=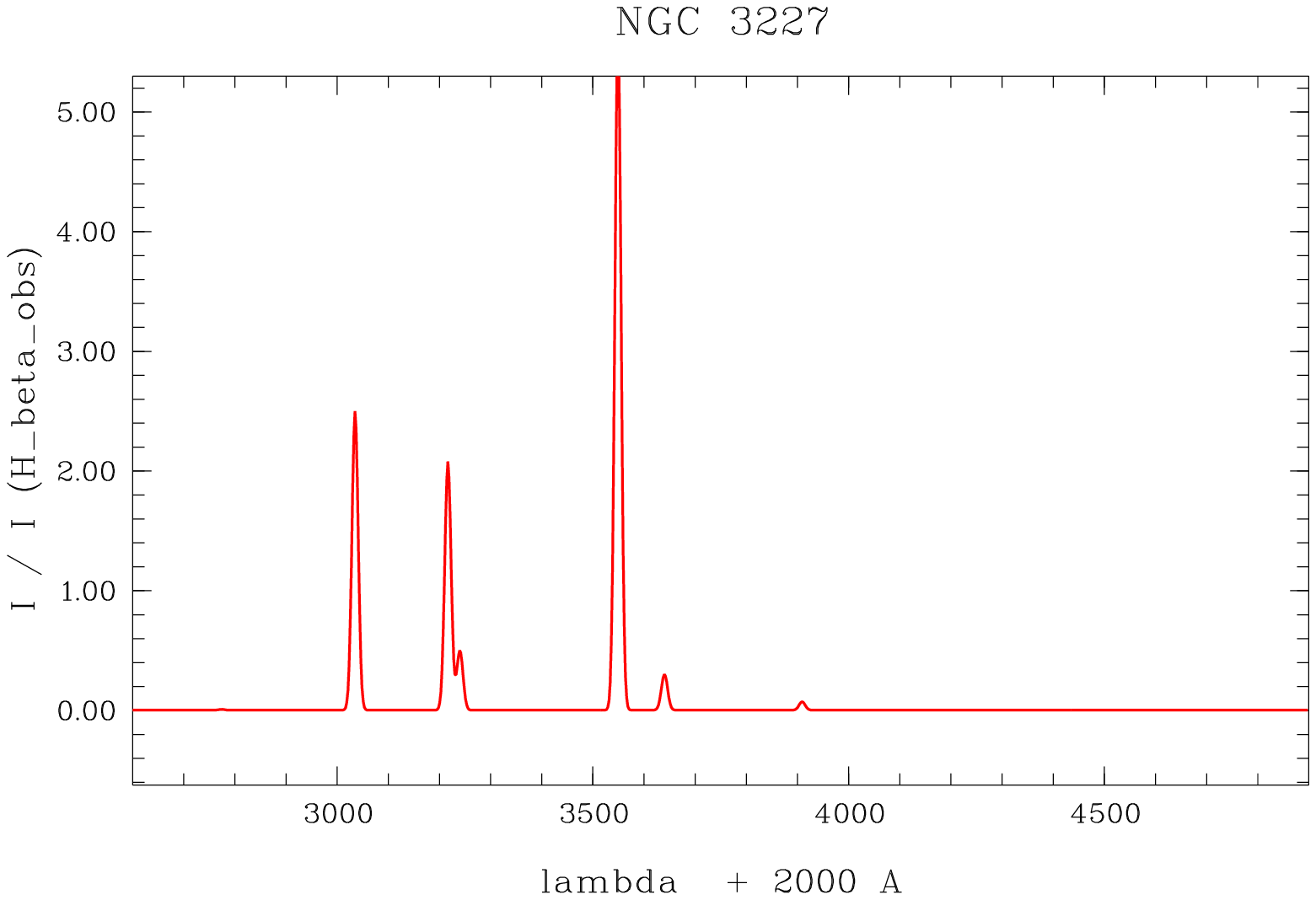,width=6.0cm,height=4.0cm,%
          bbllx=2.3cm,bblly=1.4cm,bburx=18.4cm,bbury=12.3cm,clip=}}\par
       \vspace*{-4.0cm}\hspace*{6.2cm}
      \vbox{\psfig{figure=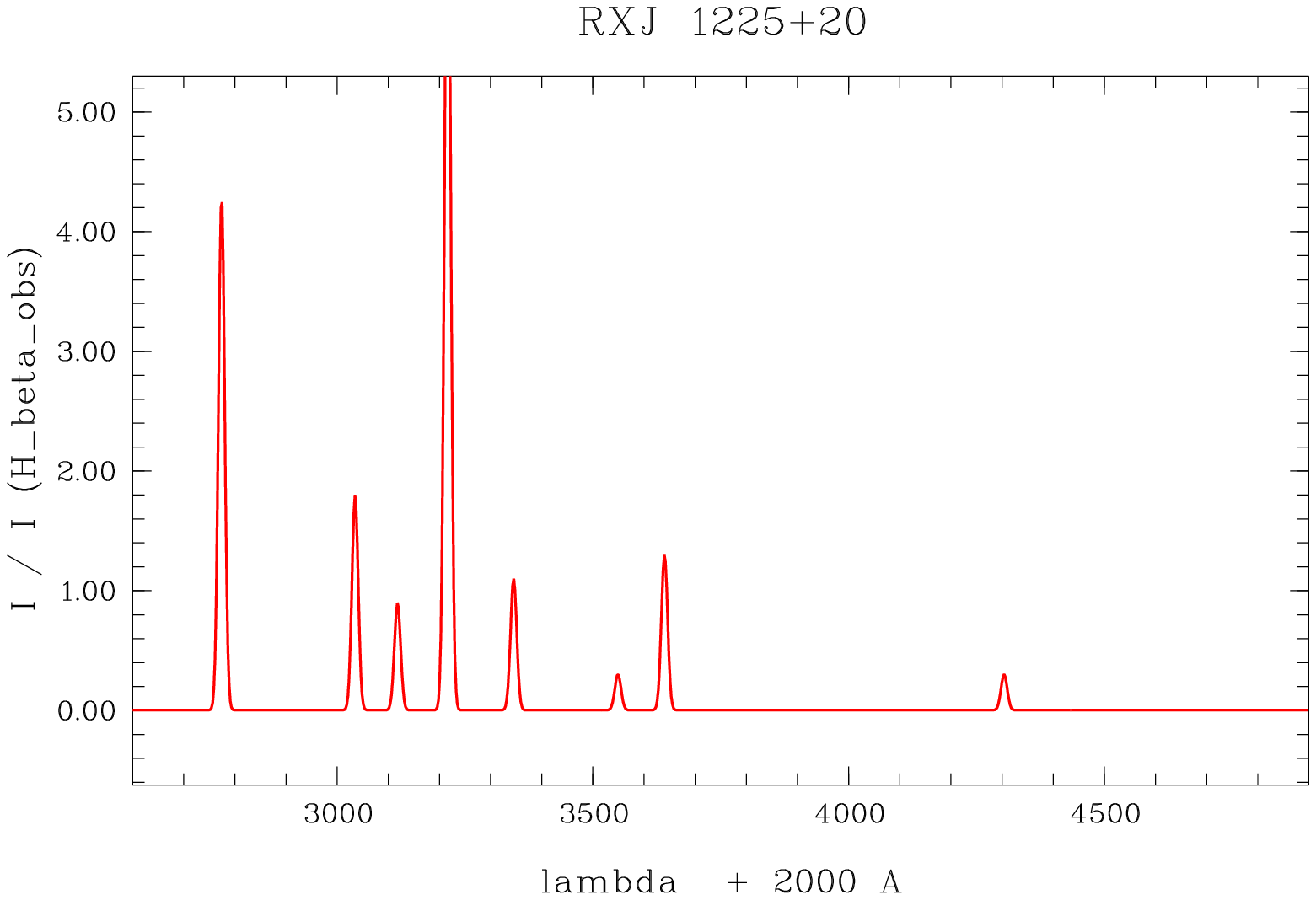,width=6.0cm,height=4.0cm,%
          bbllx=2.3cm,bblly=1.4cm,bburx=18.4cm,bbury=12.3cm,clip=}}\par
      \caption[]{Emission line spectra in the range 600 -- 2900 \AA~
predicted to arise from the warm material in individual objects compared to a mean BLR spectrum, 
to give an impression on the strengths of these lines and allow a judgement of the detectability.
The y-axis gives  
the intensity of the absorber-intrinsic emission lines relative
to the observed broad H$\beta$ emission for the individual objects.
}
      \label{emi} 
\end{figure*}

The absorber-intrinsic H$\beta$ emission
is rather large and corresponds to about 1/10 and 1/7 of the observed
L$_{\rm H\beta}$ for \144 and \45, respectively.
Scaling the predicted [FeXIV]$\lambda$5303 emission of the warm material
in order not to conflict with the observed upper limit ($ \approxlt 0.15$)
constrains the covering factor of the gas to $\approxlt$ 1/6 in \45 and
is consistent with 1 in \144.
Given the high discovery rate of supersoft AGN among X-ray selected ones
(e.g. Greiner et al. 1996),
the covering factors indeed have to be high to account for this fact.
For both objects, several strong ultraviolet emission lines are predicted, e.g.
HeII$\lambda$1640$_{\rm{wa}}$/H$\beta_{\rm{obs}}$ $\simeq 1.3$,
FeXXI$\lambda$1345$_{\rm{wa}}$/H$\beta_{\rm{obs}} \simeq 1.1$ in \144.
The corresponding emission-line spectrum is illustrated in Fig. \ref{emi} assuming the maximal
covering consistent with the data. 

\subsection{RX\,J0119.6-2821}
\glossary{RX\,J0119.6-2821}
\glossary{1E 0117.2-2837}  
\7 (1E 0117.2-2837) was discovered as an X-ray source by \ein and is
 at a redshift of $z$=0.347 (Stocke et al. 1991).  
It is serendipituously located in one of the \ros PSPC pointings. 

The X-ray spectrum is very steep. When described by a single powerlaw 
continuum with Galactic cold column, the photon index is \G $\simeq -3.6$
(--4.3, if $N_{\rm H}$ is a free parameter).
The overall quality of the fit is good ($\chi{^{2}}_{\rm red} \simeq 1$), but there are 
slight systematic
residuals around the location of absorption edges.  
Again, a successful alternative description is a warm-absorbed flat powerlaw
of canonical index.  
We find a very large column density $N_{\rm w}$ in this case, and the contribution 
of emission and reflection is no longer negligible; there is also some 
contribution to Fe K$\alpha$. For the pure absorption model, the best-fit values
for ionization parameter and warm column density are $\log U \simeq 0.8$,
$\log N_{\rm w} \simeq 23.6$ ($N_{\rm H}$ is consistent with the  Galactic value),
with $\chi{^{2}}_{\rm red}$ = 0.74. Including the contribution of emission
and reflection for 50\% covering of the warm material gives
$\log N_{\rm w} \simeq 23.8$ ($\chi{^{2}}_{\rm red}$ = 0.65).    

Several strong EUV emission lines arise from the warm material.   
Some of these are: 
FeXXI$\lambda$2304/H$\beta$ = 10, HeII$\lambda$1640/H$\beta$ = 16,  
FeXXI$\lambda$1354/H$\beta$ = 37,        
FeXVIII$\lambda$975/H$\beta$ = 16, NeVIII$\lambda$774/H$\beta$ = 9, and, just outside the
IUE sensitivity range for the given $z$, FeXXII$\lambda$846/H$\beta$ = 113. 
No absorption from CIV and NV is expected to show up. Both elements are more highly
ionized. 

\begin{figure*} 
      \vbox{\psfig{figure=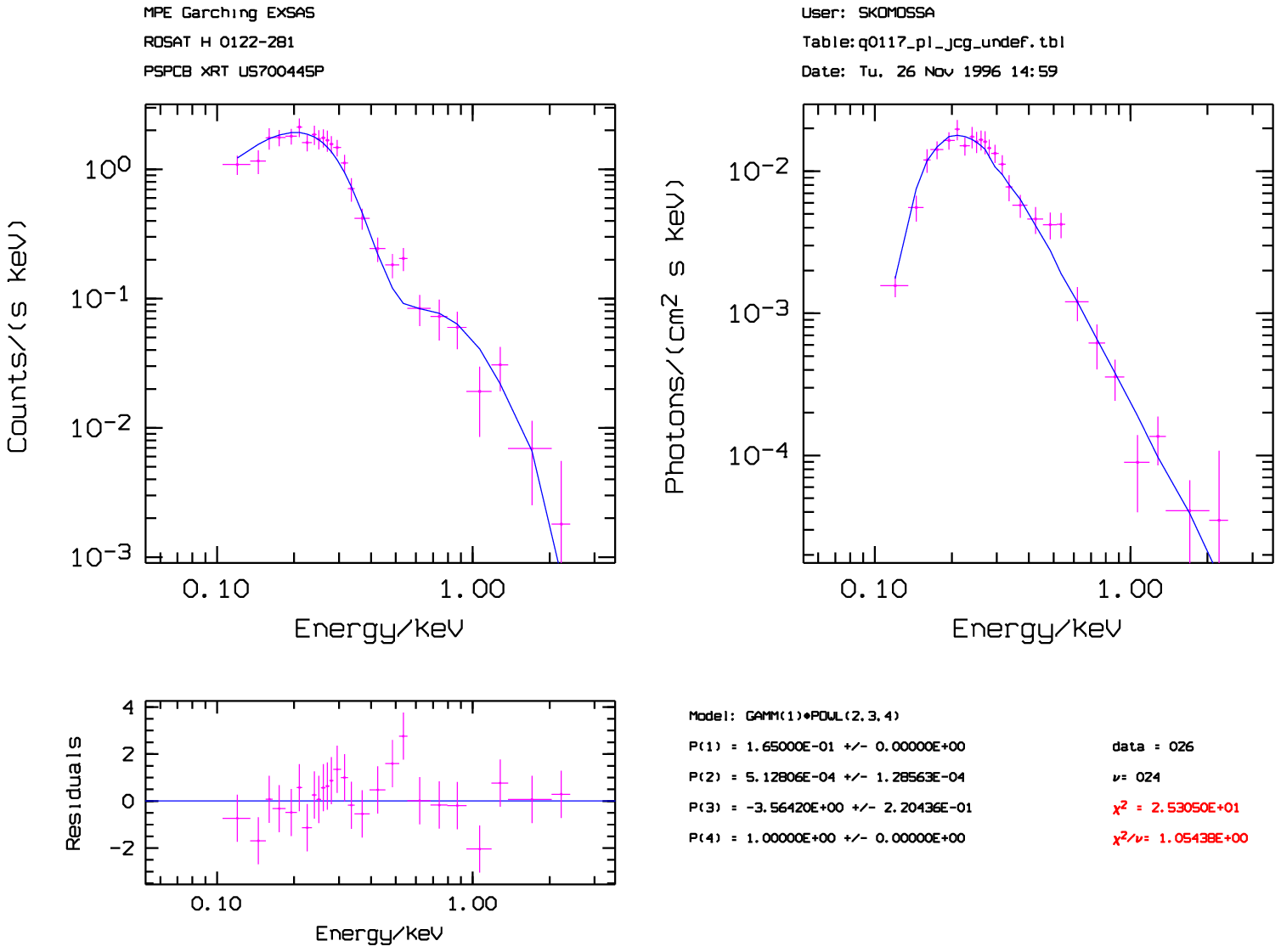,width=6.0cm,height=7.50cm,%
          bbllx=2.5cm,bblly=1.1cm,bburx=10.1cm,bbury=11.7cm,clip=}}\par
       \vspace*{-7.50cm}\hspace*{6.2cm}
      \vbox{\psfig{figure=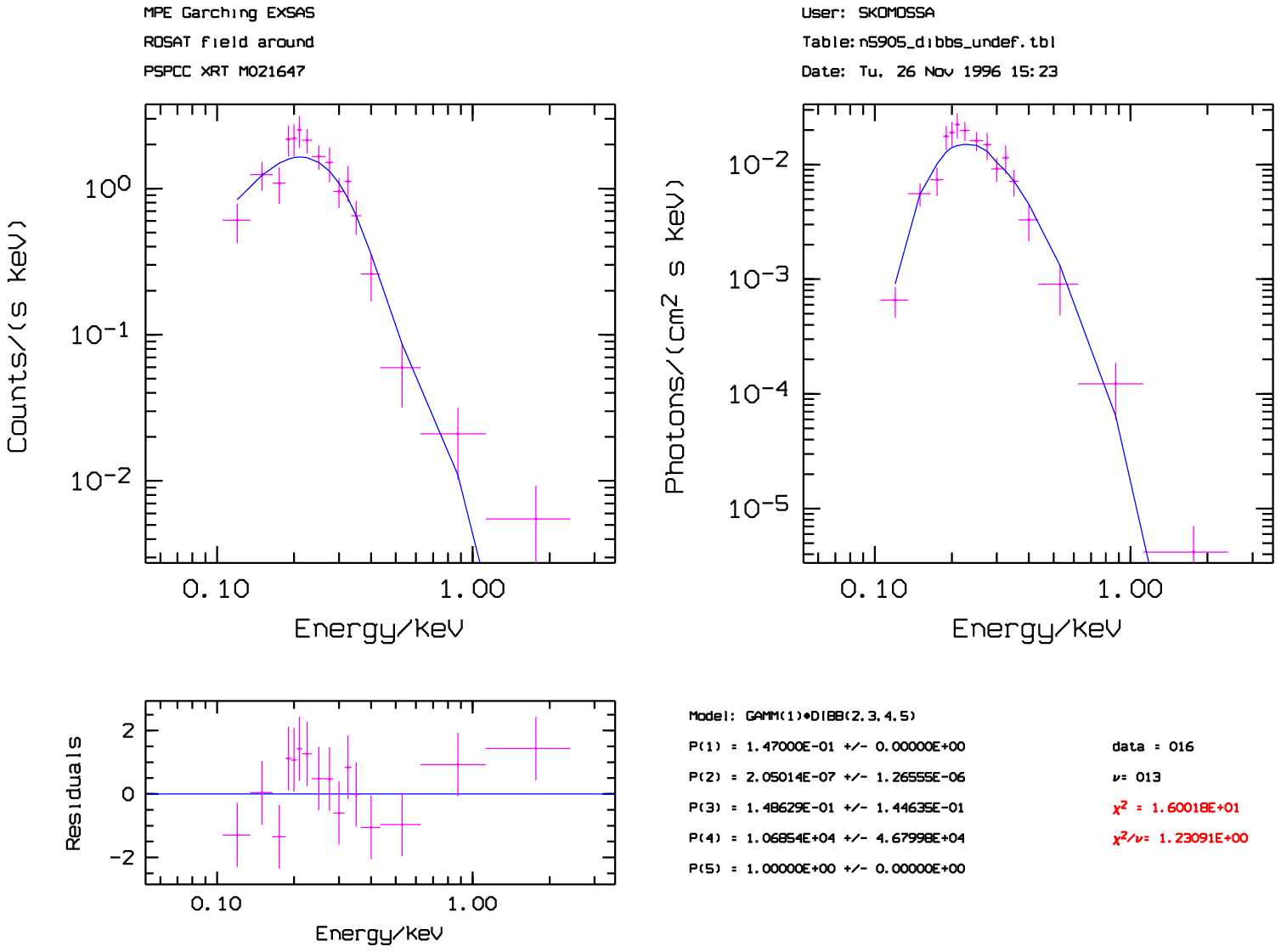,width=6.0cm,height=7.50cm,%
          bbllx=2.5cm,bblly=1.1cm,bburx=10.1cm,bbury=11.7cm,clip=}}\par
       \vspace{-0.5cm}
      \vbox{\psfig{figure=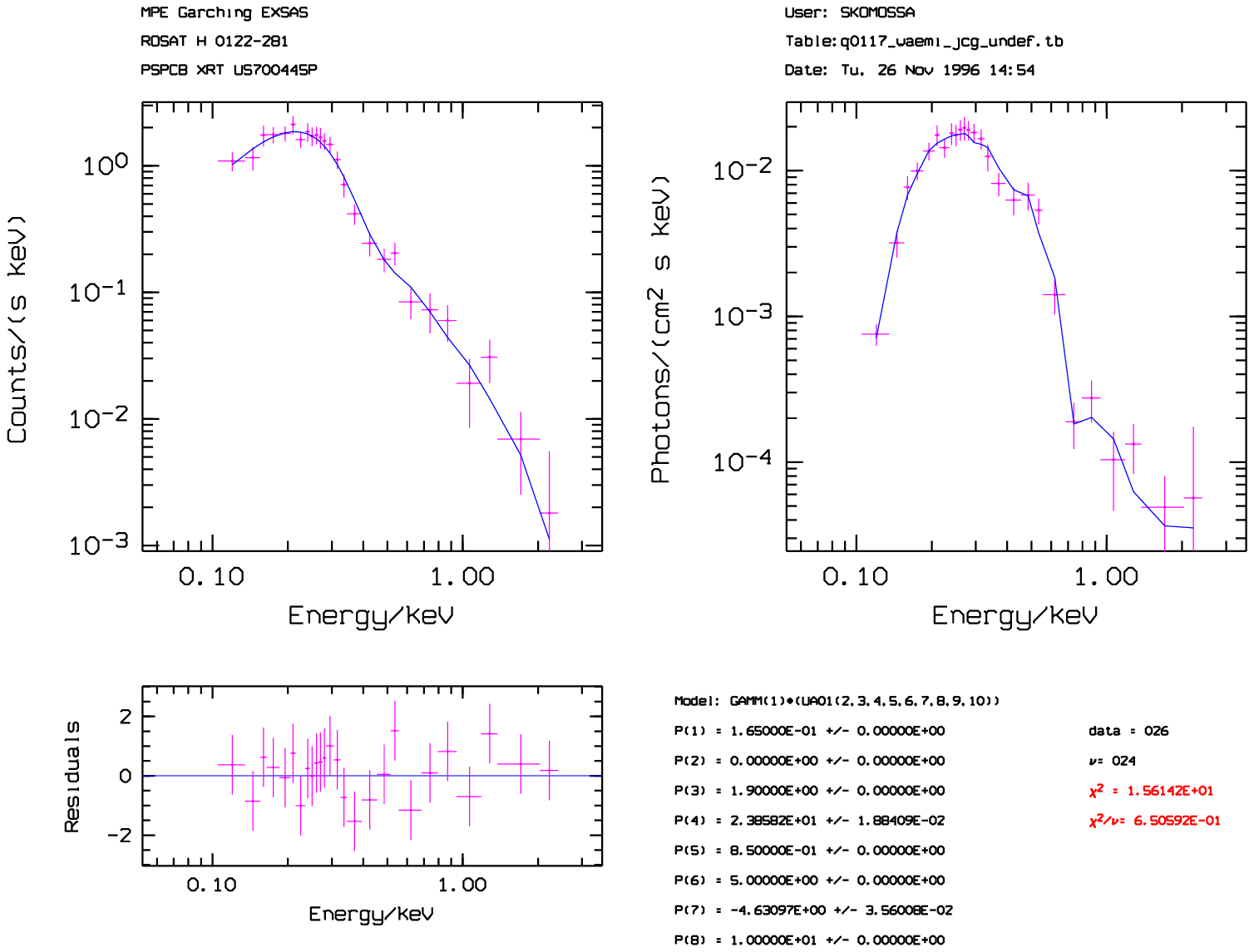,width=6.0cm,height=2.5cm,%
          bbllx=2.5cm,bblly=1.1cm,bburx=10.1cm,bbury=4.5cm,clip=}}\par
       \vspace*{-2.5cm}\hspace*{6.2cm}
      \vbox{\psfig{figure=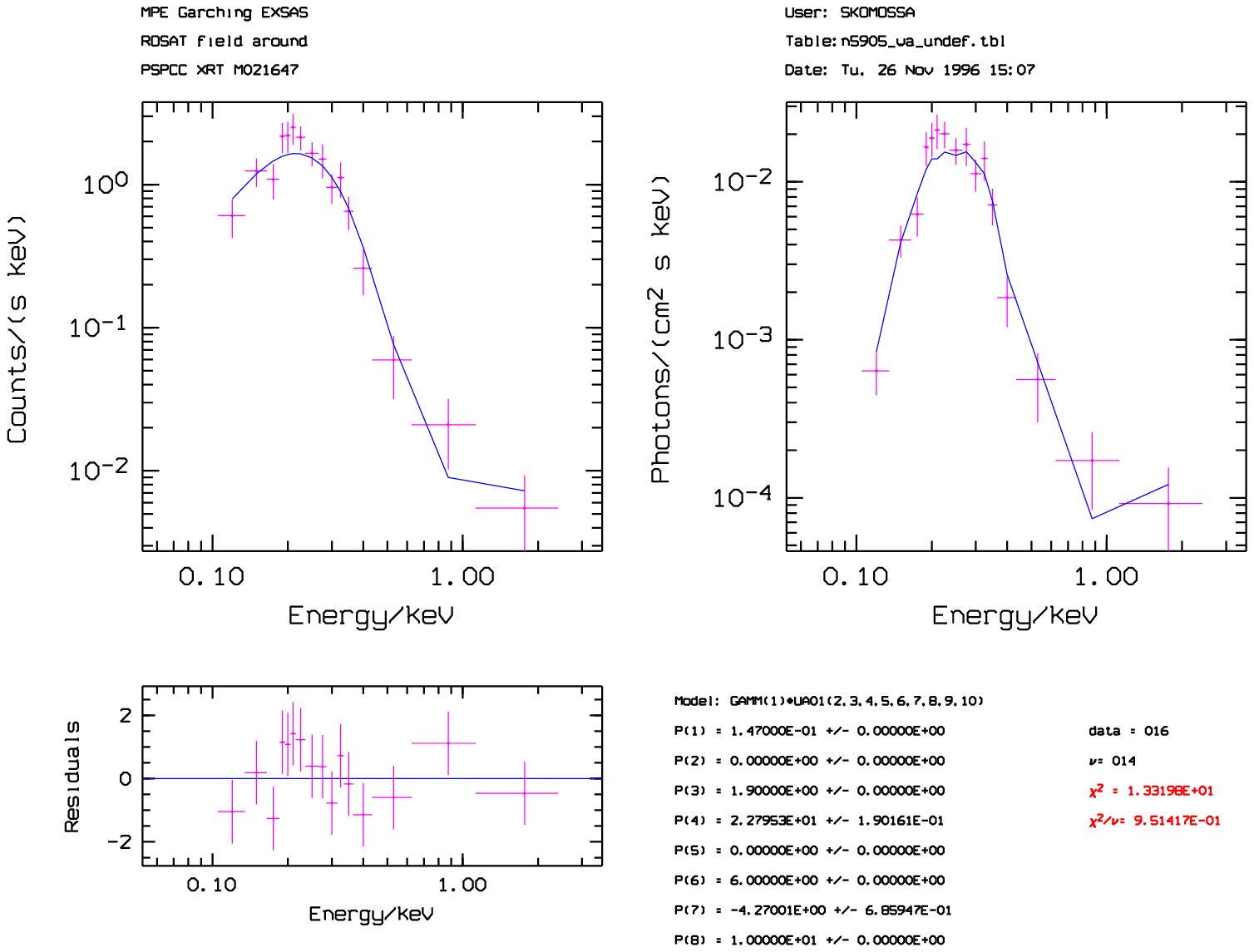,width=6.0cm,height=2.5cm,%
          bbllx=2.5cm,bblly=1.1cm,bburx=10.1cm,bbury=4.5cm,clip=}}\par
      \caption[]{Left: Powerlaw fit to the X-ray spectrum of \7 (upper panel) and residuals,
 compared to the residuals for the warm absorber fit (lower panel; note the different scale in the ordinate).
 Right: Accretion disk fit to the `outburst' observation of NGC 5905 (upper panel) and
residuals, compared to residuals for the warm absorber fit (lower panel).  
}
      \label{SED2x}
\end{figure*}

\subsection{RX\,J0134.3-4258}
\glossary{RX\,J0134.3-4258} 
This object exhibits an ultrasoft spectrum in the \ros survey data (Dec. 1990;
formally \G $\simeq -4.4$). Interestingly, the spectrum has changed to flat in
a subsequent pointing (Dec. 1992; \G $\simeq -2.2$).
This kind of spectral variability is naturally predicted within the framework of
warm absorbers, making the object a very good candidate.

\begin{figure*} 
      \vbox{\psfig{figure=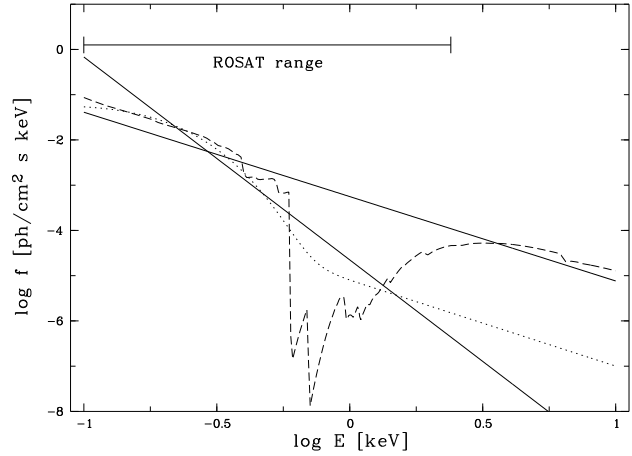,width=9.8cm,%
          bbllx=0.5cm,bblly=1.1cm,bburx=18.3cm,bbury=12.2cm,clip=}}\par
\vspace{-0.2cm} 
  \caption[]{Comparison of different X-ray spectral fits to the \ros survey
observation of \0134 (thick lines) and the pointed observation (thin solid line).
Thick solid: single powerlaw with $\Gamma_{\rm x} = -4.4$; dashed: warm-absorbed
flat powerlaw; dotted: powerlaw plus soft excess, parameterized as a black body.
All models are corrected for cold absorption. 
}
  \label{fits_0134}
\end{figure*}

We find that a warm-absorbed, intrinsically flat powerlaw can indeed describe the
survey observation.
Due to the low number of available photons, a range of possible combinations
of $U$ and $N_{\rm w}$ explains the data
with comparable success. 
A large
column density $N_{\rm w}$ (of the order 10$^{23}$ cm$^{-2}$) is needed to
account for the ultrasoft observed spectrum.
The most suggestive scenario within the framework of warm absorbers
is a change in the {\em ionization state} of ionized material along the line of sight,
caused by {\em varying irradiation} by a central ionizing source.
In the simplest case, lower intrinsic luminosity would be expected, in order to cause the deeper
observed absorption in 1990. However, the source is somewhat brighter in the survey observation
(Fig. \ref{fits_0134}).  Some variability seems to be usual, though, and the count rate changes by
about a factor of 2 during the pointed observation. If one whishes to keep this
scenario, one has to assume that the ionization state of the absorber still reflects
a preceding (unobserved) low-state in intrinsic flux.
Alternatively, gas heated by the central continuum source
may have {\em crossed the line of sight},
producing the steep survey spectrum, and has (nearly) disappeared in the 1992 observation.

It is interesting to note, that the optical spectrum of \0134 rises to the blue
(Grupe 1996),
and a soft excess on top of a powerlaw continuum may represent an
alternative explanation of the data. 

\begin{table}[ht]
\vspace{-1.5cm} 
  \caption{Comparison of different spectral fits to the NLSy1 galaxies: (i) single powerlaw (pl),
  (ii) accretion disk after Shakura \& Sunyaev (1973), and (iii) warm absorber.
  \G was fixed to --1.9 in (ii) and (iii),
  except for \0134, where \G = --2.2 (see text). 
 } 
  \begin{tabular}{llllllllll}
  \noalign{\smallskip}
  \hline
  \noalign{\smallskip}
   name &  \multicolumn{3}{l}{powerlaw$^{(1)}$~~~~~~~~~~~~} & 
                       \multicolumn{3}{l}{acc. disk + pl$^{(2)}$~~~~~~} &
                         \multicolumn{3}{l}{warm absorber$^{(2)}$} \\ 
  \noalign{\smallskip}
  \hline
  \noalign{\smallskip}
         & $N_{\rm H}^{(3)}$~~ & \G~~~~ & $\chi^2_{\rm red}$ & $M_{\rm BH}^{(4)}$~~ & 
                                       ${\dot M}\over{{\dot M_{\rm edd}}}$~ &
                                       $\chi^2_{\rm red}$ &
                                       log $U$~~ & log $N_{\rm w}$ & $\chi^2_{\rm red}$ \\
  \noalign{\smallskip}
  \hline
  \noalign{\smallskip}
    RX\,J1239.3+2431 & 0.32 & --4.3 & 1.2 & ~0.7 & 0.4 & 1.4 & --0.1 & 22.8 & 1.0 \\  
  \noalign{\smallskip}
  \hline
  \noalign{\smallskip}
    RX\,J1225.7+2055 & 0.29 & --3.7 & 1.9 & ~1.0 & 0.5 & 2.1 & ~~0.8 & 23.2 & 1.8 \\ 
  \noalign{\smallskip}
  \hline
  \noalign{\smallskip}
    RX\,J0119.6--2821 & 0.30 & --4.3 & 0.8 & ~0.6 & 0.6 & 0.7 & ~~0.8 & 23.6 & 0.7 \\ 
  \noalign{\smallskip}
  \hline
  \noalign{\smallskip}
    RX\,J0134.3--4258$^{(5)}$~~ & 0.16 & --4.4 & 0.5 & 11.4 & 0.1 & 0.6 &~~0.5 & 23.1 & 0.6 \\
  \noalign{\smallskip}
  \hline
  \noalign{\smallskip}
     \end{tabular}
  \label{tab2}

  \noindent{\small
  $^{(1)}$$N_{\rm H}$ free, if $>$ $N_{\rm H}^{\rm Gal}$ ~ $^{(2)}$$N_{\rm H}$ 
fixed to $N_{\rm H}^{\rm Gal}$ ~ $^{(3)}$in 10$^{21}$cm$^{-2}$ ~ $^{(4)}$in 10$^4$M$_{\odot}$~
$^{(5)}$survey obs.} 
\end{table}
\subsection{Warm absorbers in {\em all} NLSy1 galaxies ?} 

There are some NLSy1 galaxies in which the warm absorber 
is clearly present and causes indeed at least most of the observed 
spectral steepness, as e.g. in NGC 4051.      
Mrk 1298 is another example. It exhibits all characteristics of NLSy1s, 
except that its observed FWHM of H$\beta$, 2200 km/s, just escapes 
the criterion of Goodrich (1989).   
One of the comfortable properties of the warm-absorber interpretation
is the presence of enough photons to account for the strong observed 
FeII emission in NLSy1s, due to the intrinsically flat powerlaw.   
It does not immediately explain the occasionally observed trend of {\em stronger} FeII
in objects with {\em steeper} X-ray spectra, but it is interesting to note 
that Wang et al. (1996) find a trend for stronger FeII to preferentially occur in objects 
whose X-ray spectra show the presence of absorption edges.    

There are, however, other objects where neither a flat powerlaw with soft excess
nor a warm absorber can account for most of the spectral steepness, although
the presence of such a component cannot be excluded, as we have verified 
for {\bf 1ZwI} and {\bf PHL 1092}.  

In any model -- warm absorber, accretion-produced soft excess, or intrinsically
steep powerlaw spectrum -- it is still difficult to elegantly account for all observational trends;  
as are, besides strong FeII, the tendency for narrower FWHM of broad H$\beta$ for
steeper X-ray spectra (e.g. Boller et al., their Fig. 8), and a slight 
anti-correlation of optical and soft X-ray spectral index (Grupe 1996). 
Progress is reported in Wandel (1996). Critical comments on the distinction
between different models concerning photoionization aspects are given in
Komossa \& Greiner (1995). 

To conclude, a warm absorber was shown to successfully reproduce the 
X-ray spectra of several NLSy1s and thus may well mimic the presence of a soft excess.  
High quality spectra are 
needed to distinguish between several possible models. In fact, although the warm
absorber dominates the X-ray spectrum of the well-studied galaxy NGC 4051, an additional soft excess is
present in flux high-states, and the underlying powerlaw is steeper than usual (Sect. 3.1). This hints
to the complexity of NLSy1 spectra, with probably more than
one mechanism at work to cause the X-ray spectral steepness.         
 
\section{The X-ray outburst in the HII galaxy NGC 5905}

\glossary{NGC 5905} 
NGC 5905 underwent an X-ray `outburst' during the \ros survey observation,
with a factor $\sim$ 100 change in count rate.
The high-state spectrum is simultaneously very soft and luminous (Bade, Komossa \& Dahlem 1996). 
Its classification (based on an optical pre-outburst spectrum) is HII type.
Comparable variability was found in two further objects
(IC3599; Brandt et al. 1995, Grupe et al. 1995a, WPVS007; Grupe et al. 1995b).
The outburst mechanism is still unclear; 
tidal disruption of a star by a central black hole has been proposed.
In the following, we comment on this and other possibilities and then
discuss
whether the X-ray outburst spectrum of NGC 5905
can be explained in terms of warm absorption.

\vspace{-0.4cm}
\subsubsection {SN in dense medium}
The possibility of `buried' supernovae (SN) in {\em dense} molecular gas 
was studied by Shull (1980) and
Wheeler et al. (1980). 
In this scenario, X-ray emission originates from the shock, produced by the expansion of the SN ejecta  
into the ambient interstellar gas of high density. 
Since high luminosities can be reached this way, and the evolutionary time
is considerably speeded up, an SN in a dense
medium may be an explanation for the observed X-ray outburst in NGC 5905.
 
Assuming the observed $L_{\rm x} \approx 3 \times 10^{42}$ erg/s of NGC 5905 to be the peak luminosity,
results in a density of the ambient medium of $n_4 \simeq  \times 10^6 \rm {cm}^{-3}$
(using the estimates from Shull (1980) and Wheeler et al. (1980), and 
assuming line cooling to dominate with a cooling function of $\Lambda \propto T^{-0.6}$),
but is inconsistent with the observed softness of the spectrum: 
The expected temperature is $T \approx 10^8$ K, compared to observed one of 
$T \approx 10^6$ K.   
Additionally, fine-tuning in the column density of the surrounding medium
would be needed in order to prevent the SNR from being completely self-absorbed.
\vspace{-0.5cm}
\subsubsection{Tidal disruption of a star} 
The idea of tidal disruption of stars by a supermassive black hole (SMBH)
was originally studied as a possibility to fuel AGN, but dismissed. 
Later, Rees (1988, 1990) proposed to use individual such events 
as tracers of SMBHs in nearby galaxies.   
The debris of the disrupted star is accreted by the BH.
This produces a flare, lasting of the order
of months, with the peak luminosity in the EUV or soft X-ray spectral region.   

The luminosity emitted if the BH is accreting at its Eddington luminosity
can be estimated by $ L_{\rm edd} \simeq 1.3 \times 10^{38} M/M_{\odot}$ erg/s.
In case of NGC 5905, a BH mass of $\sim 10^4$ M$_{\odot}$ would be necessary to
produce the observed $L_{\rm x}$ (assuming it to be observed near its peak value).   
In order to reach this luminosity, a mass supply of about 1/2000 M$_{\odot}$/y
would be sufficient, assuming $\eta$=0.1.

\vspace{-0.5cm}
\subsubsection{Accretion disk instabilities}
If a massive BH exists in NGC 5905, it has to usually accrete with low
accretion rate or radiate with low efficiency, to account for the 
comparatively low X-ray luminosity of NGC 5905 in `quiescence'.
An accretion disk instability may provide an alternative explanation for the observed X-ray outburst.
Thermally unstable slim accretion disks were studied by Honma et al. (1991),
who find the disk to exhibit burst-like oscillations for the case of the standard $\alpha$
viscosity description and for certain values of accretion rate.

Using the estimate for the duration of the  high-luminosity state 
(Honma et al. 1991; their eq. 4.8), 
and assuming the duration of the outburst to be less than 5 months 
(the time difference between the first two observations of NGC 5905), 
a central black hole of mass in the range $\sim 10^4 - 10^5 M_{\odot}$
can account for the observations.
The burst-like oscillations are found by Honma et al. only for certain values of the initial accretion
rate. 
A detailed quantitative comparison with the observed outburst in NGC 5905 is difficult,
since the behavior of the disk is quite model dependent, and further detailed modeling
would be needed. 

\vspace{-0.5cm}
\subsubsection{Warm-absorbed hidden Seyfert}  
Since all previous scenarios are either unlikely (SN in dense medium) or
require some sort of fine-tuning, we finally asses the possibility of a warm-absorbed
Seyfert nucleus. 
In this
scenario, a hidden Seyfert resides within the HII-type
 galaxy, that is slightly variable and usually hidden by a cold column of absorbing
gas. The nucleus gets `visible' during its flux high-states by ionizing the originally cold column
of surrounding gas that becomes a {\em warm} absorber, thereby also accounting for
the softness of the outburst spectrum.

We find that the survey spectrum can be well described by warm absorption
with log $U \simeq$ 0.0 and log $N_{\rm w} \simeq$ 22.8.
A source-intrinsic change in luminosity by a factor of less than 10 is needed
to change the absorption to complete within the \ros band.
Variability of such order is not unusual in low-luminosity AGN; e.g., NGC 4051
has been found to be variable by about a factor of 30 within 2 years 
(Sect. 3.2).
Since there is no evidence for Seyfert activity in the optical spectrum,
the nucleus must be hidden.
Mixing dust with the warm gas
results in an optical extinction of $A_{\rm v} \approx 34^{m}$ (assuming a Galactic gas-to-dust ratio)
and would hide the Seyfert nucleus completely.
The spectrum in the
low-state  \mbox{-- with} a luminosity of $L_{\rm x} \approx 4 \times 10^{40}$ erg/s and
a shape that can be described by a powerlaw with \G $\approx 2.4$ --
can be accounted for by the usual X-ray emission of the
host galaxy (e.g. Fabbiano 1989).
However, dust with Galactic ISM properties internal to the warm gas
was shown to strongly influence the X-ray absorption structure. 
An acceptable fit to the X-ray spectrum can only be achieved by tuning the
dust properties and the dust depletion factor.

There are several ways to decide upon different scenarios:
A hidden Seyfert nucleus should reveal its presence
by a permanent hard X-ray spectral component, as observable e.g. with \asca.
One may also expect repeated flaring activity in the accretion-disk and 
warm-absorber scenario, and a \ros HRI
monitoring (PI: N. Bade) is underway.

\vspace{.2cm}\noindent{\it Acknowledgement:}
The \ros project is supported by the German Bundes\-mini\-ste\-rium
f\"ur Bildung, Wissenschaft, Forschung und Technologie (BMBF/DARA) and the Max-Planck-Society.
We thank Gary Ferland for providing {\em{Cloudy}}.


\end{document}